\documentclass[%
 reprint,
 amsmath,amssymb,
 aps,
]{revtex4-1}

\usepackage{graphicx}
\usepackage{dcolumn}
\usepackage{bm}
\usepackage{longtable}
\usepackage{enumerate}
\usepackage{subfigure}
\usepackage{multirow}
\usepackage{cancel}
\usepackage{tabularx}
\usepackage{threeparttable}
\usepackage{dcolumn}
\usepackage{amssymb}
\usepackage{afterpage}
\usepackage{nicefrac}

\usepackage{ulem}
\newcommand{\down}{\sout{$\downarrow$}}
\newcommand{\up}{\sout{$\uparrow$}}
\newcommand{\zero}{\sout{\phantom{$\downarrow \negthickspace \uparrow$}}}
\newcommand{\double}{\sout{$\downarrow \negthickspace \uparrow$}}

\usepackage[version=3]{mhchem}
\usepackage{braket}
\usepackage{amsmath}

\begin{document}


\title{On the Multi-Reference Nature of Plutonium Oxides: \ce{PuO2^{2+}}, \ce{PuO2}, \ce{PuO3} and \ce{PuO2(OH)2}}

\author{Katharina Boguslawski}
\email{k.boguslawski@fizyka.umk.pl}
\affiliation{%
 Institute of Physics, Faculty of Physics, Astronomy and Informatics, Nicolaus Copernicus University in Torun, Grudzi{a}dzka 5, 87-100 Toru{n}, Poland}
\author{Florent R\'{e}al}
\affiliation{
Univ. Lille, CNRS, UMR 8523 - PhLAM - Physique des Lasers Atomes et Mol{\'e}cules, F-59000 Lille, France
}
\author{Pawe{\l} Tecmer}
\email{ptecmer@fizyka.umk.pl}
\affiliation{
 Institute of Physics, Faculty of Physics, Astronomy and Informatics, Nicolaus Copernicus University in Torun, Grudzi{a}dzka 5, 87-100 Toru{n}, Poland}
\author{Corinne Duperrouzel}
\affiliation{
Univ. Lille, CNRS, UMR 8523 - PhLAM - Physique des Lasers Atomes et Mol{\'e}cules, F-59000 Lille, France
}
\affiliation{
Department of Chemistry and Chemical Biology, McMaster University, Hamilton, 1280 Main Street West, L8S 4M1, Canada
}
\author{Andr{\'e} Severo Pereira Gomes}
\affiliation{
Univ. Lille, CNRS, UMR 8523 - PhLAM - Physique des Lasers Atomes et Mol{\'e}cules, F-59000 Lille, France
}
\author{{\"O}rs Legeza}
\affiliation{
Strongly Correlated Systems ``Lend\"ulet" Research Group, Wigner Research Center for Physics, H-1525 Budapest, Hungary
}
\author{Paul W. Ayers}
\affiliation{
Department of Chemistry and Chemical Biology, McMaster University, Hamilton, 1280 Main Street West, L8S 4M1, Canada
}
\author{Val{\'e}rie Vallet}
\email{valerie.vallet@univ-lille1.fr}
\affiliation{
Univ. Lille, CNRS, UMR 8523 - PhLAM - Physique des Lasers Atomes et Mol{\'e}cules, F-59000 Lille, France
}

\begin{abstract}
{Actinide-containing complexes present formidable challenges for electronic structure methods due to the large number of degenerate or quasi-degenerate electronic states arising from partially occupied 5f and 6d shells.
    Conventional multi-reference methods can treat active spaces that are often at the upper limit of what is required for a proper treatment of species with complex electronic structures, leaving no room for verifying their suitability.
    In this work we address the issue of properly defining the active spaces in such calculations, and introduce a protocol to determine optimal active spaces based on the use of the Density Matrix Renormalization Group algorithm and concepts of quantum information theory. We apply the protocol to elucidate the electronic structure and bonding mechanism of volatile plutonium oxides (\ce{PuO3} and \ce{PuO2(OH)2}), species associated with nuclear safety issues for which little is known about the electronic structure and energetics. 
    We show how, within a scalar relativistic framework, orbital-pair correlations can be used to guide the definition of optimal active spaces which provide an accurate description of static/non-dynamic electron correlation, as well as to analyse the chemical bonding beyond a simple orbital model. From this bonding analysis we are able to show that the addition of oxo- or hydroxo-groups to the plutonium dioxide species considerably changes the pi-bonding mechanism with respect to the bare triatomics, resulting in \textit{bent} structures with considerable multi-reference character.
}
\end{abstract}

\maketitle
\section{Introduction}
Plutonium oxides are ubiquitous in the nuclear industry, either as components of 
nuclear fuels or of fission products\cite{Hartmann2012, Haire2001}. Due to its 
extreme radio toxicity, the consequences of an eventual release of plutonium to 
the environment in the event of nuclear (such as in the Fukushima Daiichi plant, 
where one reactor was loaded with mixed uranium and plutonium oxide (MOX) fuels) 
or industrial accidents (like a solvent fire in a facility performing the PUREX~\cite{PUREX1,PUREX2} 
process for reprocessing of spent nuclear fuel) would be severe. 

In preparing for such eventualities it is essential to be able to predict the most likely 
state of the metal, what kind of chemical reactions can potentially occur, and 
the effect of these in the dispersion of these species outside the containment vessels etc. It is known, 
for instance, that in the presence of steam and oxygen, plutonium is released primarily as 
\ce{PuO3(g)} and \ce{PuO2(OH)2(g)}, and less probably as \ce{PuO2(g)}~\cite{Ronchi-JNM2000},
meaning that these could be a potentially important source of radioactive aerosols and
transported over long distances.

The considerable difficulties in performing systematic experiments with these materials
mean that the reactivity of plutonium and other extremely radioactive species is still
not very well understood. Theoretical studies are an alternative to experiments, though
a reliable modeling of the electronic structures and chemical properties of 
plutonium-containing complexes remains a challenge for present-day quantum chemistry, as
it requires treating static and dynamical correlation effects, relativistic (scalar and
spin-orbit coupling) effects and eventually the influence of the species' immediate surroundings, 
if processes take place in solution or in the solid state. 

Progresses in four-component or two-component approaches notwithstanding, a treatment of relativistic effects via 
one-component  approaches remains computationally advantageous.~\cite{Reiher_book,Tecmer2016} 
As scalar relativistic effects are nowadays relatively straightforward to treat, and there
are sufficiently accurate approaches to treat spin-orbit interactions perturbatively~\cite{EPCISO},
a remaining difficulty is the large number of competing highly-correlated electronic states resulting from distributing 
electrons among the energetically close-lying 5f, 6d, and 7s atomic orbitals, which requires a 
multi-reference treatment.
In addition, the so-called inner valence orbitals, 6s and 6p, are highly polarizable and yield 
non-negligible contribution to the electron correlation energy.~\cite{schreckenbach_dft,gomes2015applied}
These peculiarities in the electronic structure of actinide compounds significantly increase the 
computational requirements as the number of electrons that have to be considered in the correlation 
treatment usually exceeds the current limit of conventional multi-reference methods.~\cite{Maron-pu022+-1999,Clavaguera-Sarrio_2004} 

Several density functional theory (DFT) calculations have been performed on plutonium complexes such as oxides 
in the past.~\cite{Trond_1999,Hay2000,schreckenbach_dft,U_core_potentials,Kovacs_11,Kovacs_12,Huang2013,Kovacs_15}
However, albeit DFT is in principle capable of treating open-shell molecules with multi-reference 
character,~\cite{DFT-Cremer-IJMS2002-3-604} it is known that the currently available density functional 
approximations often perform poorly in such cases.~\cite{Clavaguera-Sarrio_2004,actinide-Gomes-PCCP2014-16-9238}
It is therefore preferable to employ from the outset multi-reference approaches such as the Complete-Active-Space 
Self-Consistent-Field (CASSCF) method and the Density Matrix Renormalization Group (DMRG) 
algorithm~\cite{white,white2,white-qc,scholl05,ors_springer,marti2010b,chanreview,wouters-review,Ors_ijqc,yanai-review}
in order to obtain a qualitatively correct zeroth-order wave function for these systems.

An appealing feature of the DMRG algorithm in the case of actinides and their potentially large
number of quasi-degenerate states is that it allows us to include the full valence space of molecular 
orbitals in one single calculation without restricting the configuration space. Furthermore, 
when combined with concepts of quantum information theory, DMRG allows us to quantify orbital entanglement~\cite{DEAS}
and orbital-pair correlations~\cite{Rissler2006,Barcza_11,Barcza2013,entanglement_letter,entanglement_bonding_2013,barcza2014entanglement,Kasia_ijqc,Ors_ijqc}
that enable us to gain a better understanding of electron correlation effects,~\cite{entanglement_letter,Keller2015,Boguslawski2016} elucidate chemical bonding in molecules,~\cite{entanglement_bonding_2013,PCCP_bonding,bonding_qit,boguslawski2014chemical,Roland-RuNO,AP1roG-actinides,Zhao2015} and detect changes in the electronic wave function.~\cite{Ors-LiF-TTNS,MIT-Fertita-2014,Corinne_2015}
The suitability of DMRG for helping to understand the electronic structure of actinides can be seen
in a recent study of the changes in the ground-state for the CUO molecule when diluted
in different noble gas matrices.~\cite{CUO_DMRG}

\begin{figure}
\includegraphics[width=0.9\columnwidth]{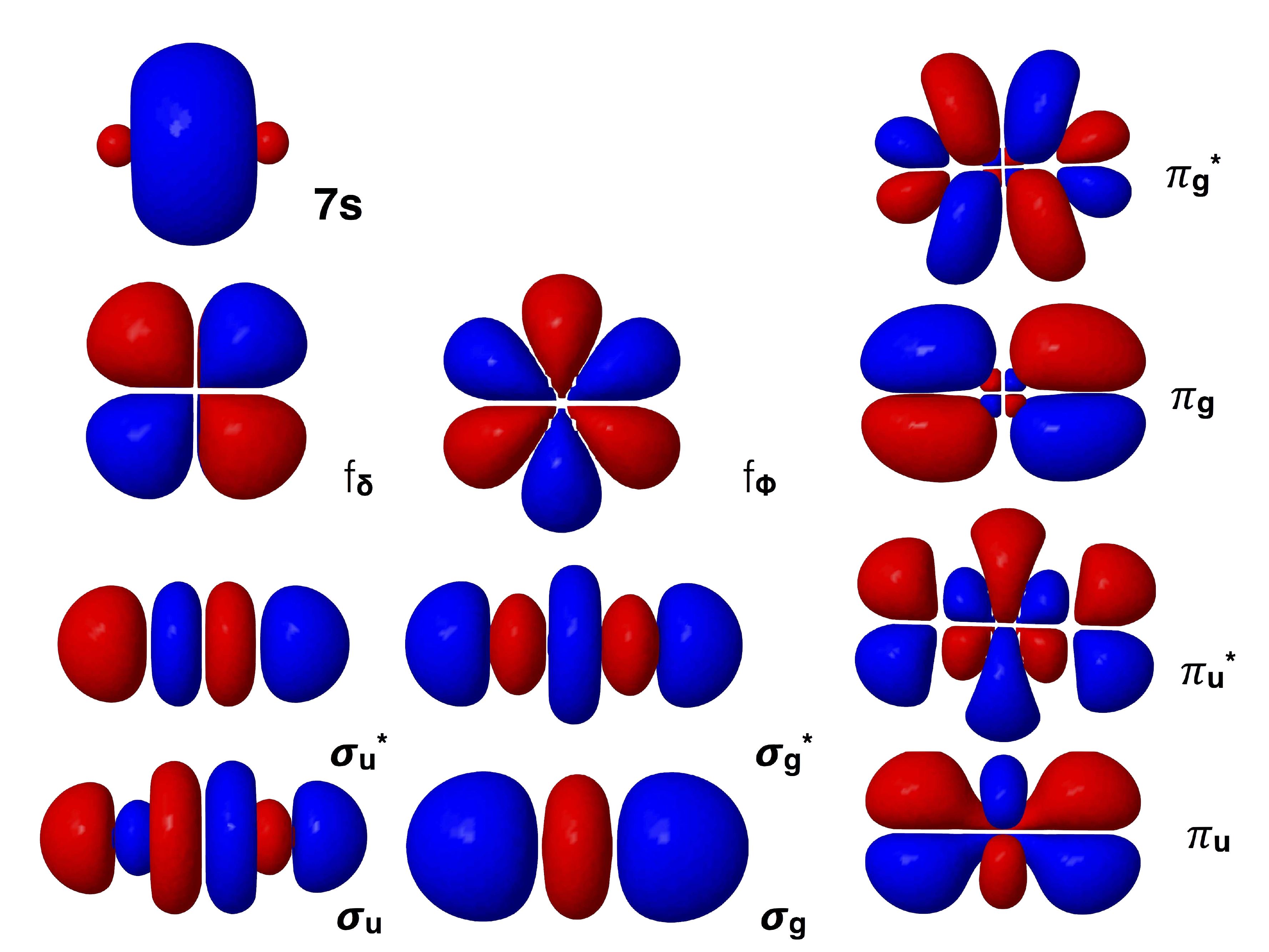}
\caption{Spin-free, bonding, antibonding, and nonbonding orbitals of actinyl molecule.}
\label{fig:orbs_PuO22}
\end{figure}

The considerable body of work on actinyl species such as \ce{PuO2^{2+}} has allowed us to be
confident in our understanding of their electronic structure and the nature of the 
actinide--oxygen bond.~\cite{Trond_1999,actinide-Maron-CP1999-244-195,Kovacs_12,fscc-npo2,ivan_puo2_08} 
Using molecular orbital diagrams and accounting for point group symmetry, plutonium and 
oxygen can form $\sigma$ and $\pi$ bonds with doubly occupied orbitals up until the 3$\sigma_u$ orbital.
With the 5f and 6d orbitals being in the primary valence shell, actinides are able to form unique bonds 
and interact with other compounds in ways that no other elements are able 
to.~\cite{denning07,Actinides_bible,actinoid_rev_2012,denning_91b,Castro-Rodriguez2004,Baker2012}
In plutonium dioxides, the 5f and 6d orbitals of the plutonium center interact with the valence 
orbitals on the oxygen atoms, with the exception of four pairwise degenerate nonbonding orbitals: 
the doubly degenerate ${\delta_u}$, ${\phi_u}$ and $\delta_g$ orbitals.~\cite{denning07}. The 
bonding, antibonding, nonbonding orbitals are represented in Figure~\ref{fig:orbs_PuO22}.
This information can also be applied to neutral actinide dioxides, such as \ce{PuO2}.\cite{ivan_puo2_08} 

However, the bonding mechanism of larger plutonium oxides and oxyhydroxides, such as \ce{PuO3} 
or \ce{PuO2(OH)2}, is still largely unknown. For the former, its optimal geometry has been computed by 
Zaitsevskii~\textit{et al.}~\cite{actinide-Zaitsevskii-DC2013-448-1} using scalar relativistic 
DFT calculations, without mentioning the nature of the electronic ground state. In an earlier 
work, Gao~\textit{et al.}~predicted a $\mathrm{^7B_1}$ ground state, also using scalar relativistic DFT 
calculations,~\cite{actinide-Gao-AC2004-62-454} but given the difficulties of DFT in treating 
the multi-reference character of a wave function, these findings must be cross-checked.

Our primary objective in this work is therefore to investigate in detail the electronic structure 
of \ce{PuO2}, \ce{PuO3}, and \ce{PuO2(OH)2} in a scalar relativistic framework and to determine the 
most important valence orbitals that should be included in the active space of currently available 
multi-reference methods, such as CASPT2 with \textit{a posteriori} spin-orbit coupling treatment, 
to ultimately compute very accurate thermodynamic properties in a forthcoming publication. 

This work is organized as follows. Section~\ref{sec:comp} summarizes the computational details.
Numerical results, including an orbital-pair correlation and entropy-based bonding analysis, are presented in section~\ref{sec:results}.
Finally, we conclude in section~\ref{sec:conclusions}.

\section{Computational Details}\label{sec:comp}

\subsection{Basis set}
All calculations used a relativistic effective core pseudo potential (RECP) for the plutonium center with the corresponding contracted basis set made of $\mathrm{(14s13p10d8f6g)} \rightarrow \mathrm{[6s6p5d4f3g]}$.~\cite{ECP-Andrae-TCA1990-77-123--141,basis-Cao-JCP2003-118-487-496, basis-Cao-JMST2004-673-203---209}
Oxygen and hydrogen atoms are described by the  aug-cc-pVTZ basis sets.\cite{basis-Dunning-JCP1989-90-1007,basis-Kendall-JCP1992-96-6796}
The contraction schemes are as follows:
O:$\mathrm{(11s6p3d2f)} \rightarrow \mathrm{[5s4p3d2f]}$,
H:$\mathrm{(6s3p2d)} \rightarrow \mathrm{[4s3p2d]}$.

\subsection{Geometry Optimization}
\begin{figure}
\includegraphics[width=0.9\columnwidth]{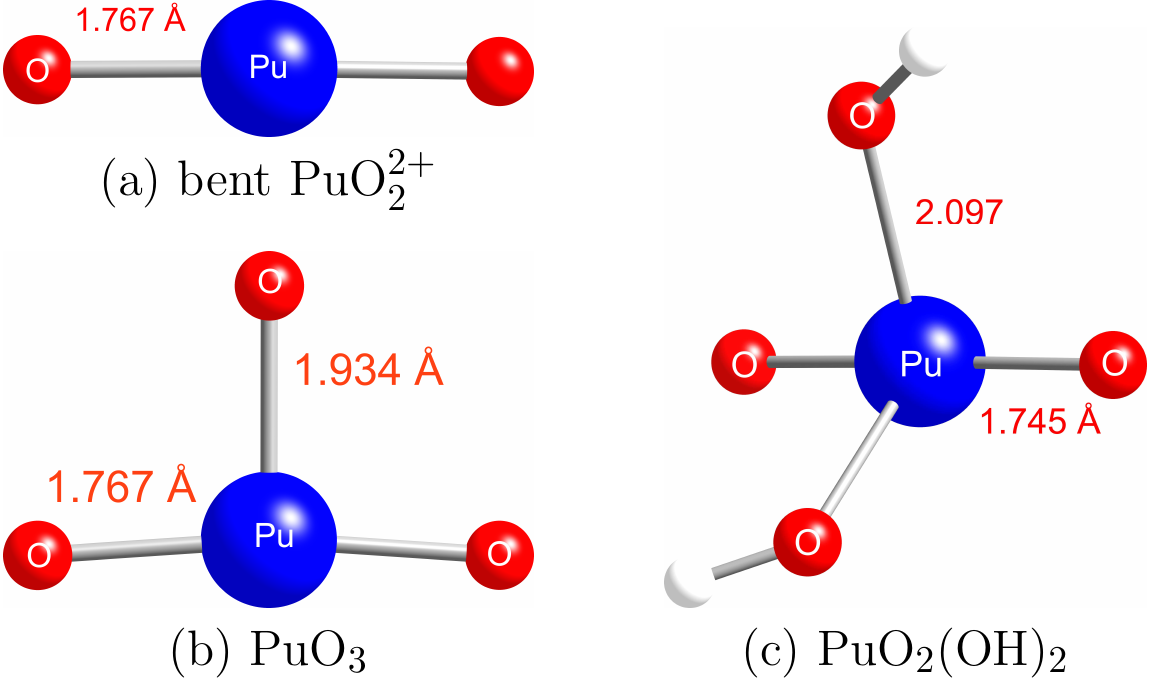}
\caption{Spin-free structures of all investigated plutonium oxides. The xyz coordinates of the DFT-optimized structures are summarized in the ESI.}
\label{fig:structures}
\end{figure}

Since the exact nature of the electronic ground-states are not known for \ce{PuO3} and \ce{PuO2(OH)2}, approximate structures of \ce{PuO2}, \ce{PuO2^{2+}}, \ce{PuO3}, and \ce{PuO2(OH)2} were optimized using the \textsc{GAUSSIAN09} software package to explore the properties of these oxides.~\cite{prog-G09}
The B3LYP hybrid exchange-correlation functional~\cite{dft-Becke-JCP1993-98-5648-5652} was used.
The structure of the \textit{bent} \ce{PuO2^{2+}} was obtained from the optimized \ce{PuO3} by removing the distant oxygen. 
The electronic structure of \ce{PuO2} was optimized for the quintet $\mathrm{A_g}$ state, while the electronic structures of all other molecules were optimized for the corresponding triplet states of $\mathrm{B2}$ and $\mathrm{B}$ symmetries, which were found to be most stable for the \ce{PuO3} and \ce{PuO2(OH)2} molecules, respectively. In the  \ce{PuO2^{2+}} and \ce{PuO2} molecules,  the optimized bond lengths are 1.711~{\AA} and  1.814~{\AA} respectively. The two bond distances are different from those computed by La Macchia~\textit{et al.}~due to the use of different basis sets.~\cite{ivan_puo2_08}
The xyz geometries of all molecules are presented in Figure~\ref{fig:structures} and available in the ESI\dag. 

\subsection{CASSCF}
All CASSCF~\cite{Roos_casscf, Werner_1985} calculations were performed using the \textsc{MOLPRO}2012~\cite{molpro2012, molpro-WIREs} software suite. 
For all plutonyl molecules, \textit{i.e.}, \ce{PuO2^{2+}} and \ce{PuO2}, the active orbital spaces consist of ${\delta}_u$, ${\phi}_u$, and $\pi_u^*$ orbitals.
Specifically, for the linear and \textit{bent} \ce{PuO2^{2+}} molecules, the active space comprises two electrons and six orbitals (CAS(2,6)SCF). The quintet ground state of the \ce{PuO2} molecule is described by distributing four electrons in six orbitals of the active space (CAS(4,6)SCF).  
$D_{2h}$ point-group symmetry was used for the linear \ce{PuO2^{2+}} and \ce{PuO2} molecules. The energetically lowest wave-functions are the two lowest triplet states of $\mathrm{B_{2g}}$ and $\mathrm{B_{3g}}$ symmetries for \ce{PuO2^{2+}}. In the case of \ce{PuO2}, the wave function is dominated by one single determinant (${\delta}_u^{(1)}$,${\delta}_u^{(1)}$, ${\phi}_u^{(1)}$, ${\phi}_u^{(1)}$ ) and of $\mathrm{A_g}$ symmetry with a significant configuration corresponding to the contribution of (${\phi}_u^{(1)}$, ${\phi}_u^{(1)}$, ${\pi}_u^{*{(1)}}$, $\pi_u^{*{(1)}}$).



For both \ce{PuO3} and \ce{PuO2(OH)2}, the active space contains two electrons and four orbitals (CAS(2,4)SCF). 
$C_{2v}$ point-group symmetry was imposed for the \textit{bent} \ce{PuO2^{2+}} and \ce{PuO3} molecules, while $C_2$ point-group symmetry was used for \ce{PuO2(OH)2}. The associated triplet ground-state wave functions are of symmetry $\mathrm{B_2}$ for \ce{PuO2^{2+}} and \ce{PuO3}, and of $\mathrm{B}$ symmetry for \ce{PuO2(OH)2}, respectively.
The CASSCF orbitals were visualized using the Jmol visualization software~\cite{Jmol} and are shown in each orbital-pair correlation diagram below.

\subsection{DMRG}
The \textsc{Budapest DMRG}~\cite{dmrg_ors} program was used to perform the DMRG calculations. The natural orbitals obtained from the CASSCF calculations as described in the previous subsection were used as the orbital basis.

For all DMRG calculations, the CASSCF active orbital spaces were extended by including additional occupied and virtual orbitals.
For the simplest molecules, \textit{i.e.}, \ce{PuO2^{2+}} and \ce{PuO2}, the original CAS was extended by adding to the ${\delta}_u$, ${\phi}_u$, and ${\pi}_u^*$ orbitals all remaining bonding and antibonding orbitals as well as the 6d$_{\delta}$ and 7s ones, resulting in a total of 25 molecular orbitals as well as 26 and 28 electrons, respectively. We will refer to these extended active spaces as the full valence CAS (FV-CAS).
Furthermore, for \textit{bent} \ce{PuO2^{2+}}, 12 occupied orbitals (5 in A$_1$, 2 in B$_1$, 4 in B$_2$ and 1 in A$_2$) and 8 additional virtual orbitals (3 in A$_1$, 2 in B$_1$, 2 in B$_2$, and 1 in A$_2$) were added to the CASSCF active space, resulting in DMRG(26,26).
For \ce{PuO3}, we included 16 occupied orbitals (7 in A$_1$, 3 in B$_1$, 5 in B$_2$ and 1 in A$_2$) and 6 virtual orbitals (3 in A$_1$, 2 in B$_1$ and 1 in B$_2$) with respect to CASSCF, increasing it to DMRG(34,26). 
The CASSCF active space of \ce{PuO2(OH)2} was extended by 20 occupied orbitals (10 in A and 10 in B) and 11 virtual orbitals (6 in A and 5 in B), yielding DMRG(42,35).
To facilitate our notation, all above-mentioned DMRG active spaces will be indicated as FV-CAS.

For each molecule, we investigated a second active orbital space that was constructed by including only the strongly correlated orbitals in the DMRG active space.
These strongly correlated orbitals were identified using the orbital entanglement and correlation measures obtained from DMRG calculations for the aforementioned extended full-valence active orbital spaces (orbital-pair correlation ${I_{i|j} > 0.01}$ as a selection threshold, \textit{vide infra}).
If the size of the resulting correlation-based active space was too big or if only one orbital was close to the cut-off threshold (as in \ce{PuO3}, bent \ce{PuO2^{2+}}, and \ce{PuO2(OH2)}), we further increased our selection criterion to a value of $I_{i|j}\approx0.02$ (\textit{vide infra}), which results in neglecting orbitals with only one major orbital-pair correlation close to the threshold of ${I_{i|j} \approx 0.01}$. 
Specifically these are
DMRG(14,16) for linear \ce{PuO2^{2+}} (2 in A$_g$, 3 in B$_{3u}$, 3 in B$_{2u}$, 3 in B$_{1u}$, 2 in B$_{2g}$, 2 in B$_{3g}$, and 1 in A$_u$),
DMRG(14,15) for \textit{bent} \ce{PuO2^{2+}} (4 in A$_1$, 3 in B$_1$, 5 in B$_2$, and 3 in A$_2$),
DMRG(18,17) for \ce{PuO2} (1 in A$_g$, 4 in B$_{3u}$, 4 in B$_{2u}$, 3 in B$_{1u}$, 2 in B$_{2g}$, 2 in B$_{3g}$, and 1 in A$_u$),
DMRG(14,14) for \ce{PuO3} (4 in A$_1$, 5 in B$_1$, 4 in B$_2$, and 1 in A$_2$),
and finally
DMRG(20,22) for \ce{PuO2(OH)2} (10 in A, 12 in B). In the following, we will abbreviate those two DMRG active spaces as optCAS, indicating the optimized DMRG active space determined by means of orbital entanglement and correlation, and FV-CAS, referring to the largest DMRG active space as mentioned in the previous paragraph.

To enhance convergence, we optimized the orbital ordering.~\cite{Barcza_11} The initial guess was generated using the dynamically extended-active-space procedure (DEAS).~\cite{DEAS}
For block states $m>512$, we used the dynamic block state selection (DBSS) approach~\cite{legeza2003,legeza2004} and set the quantum information loss $\chi=10^{-5}$ and the minimum number of block states $m_{\mathrm{min}}=512$, while the maximum number was set to $m_{\mathrm{max}}=\{1024,2048\}$. All DMRG calculations are summarized in the ESI.

\begin{figure*}[!htbp]
\centering
\includegraphics[width=0.7\textwidth]{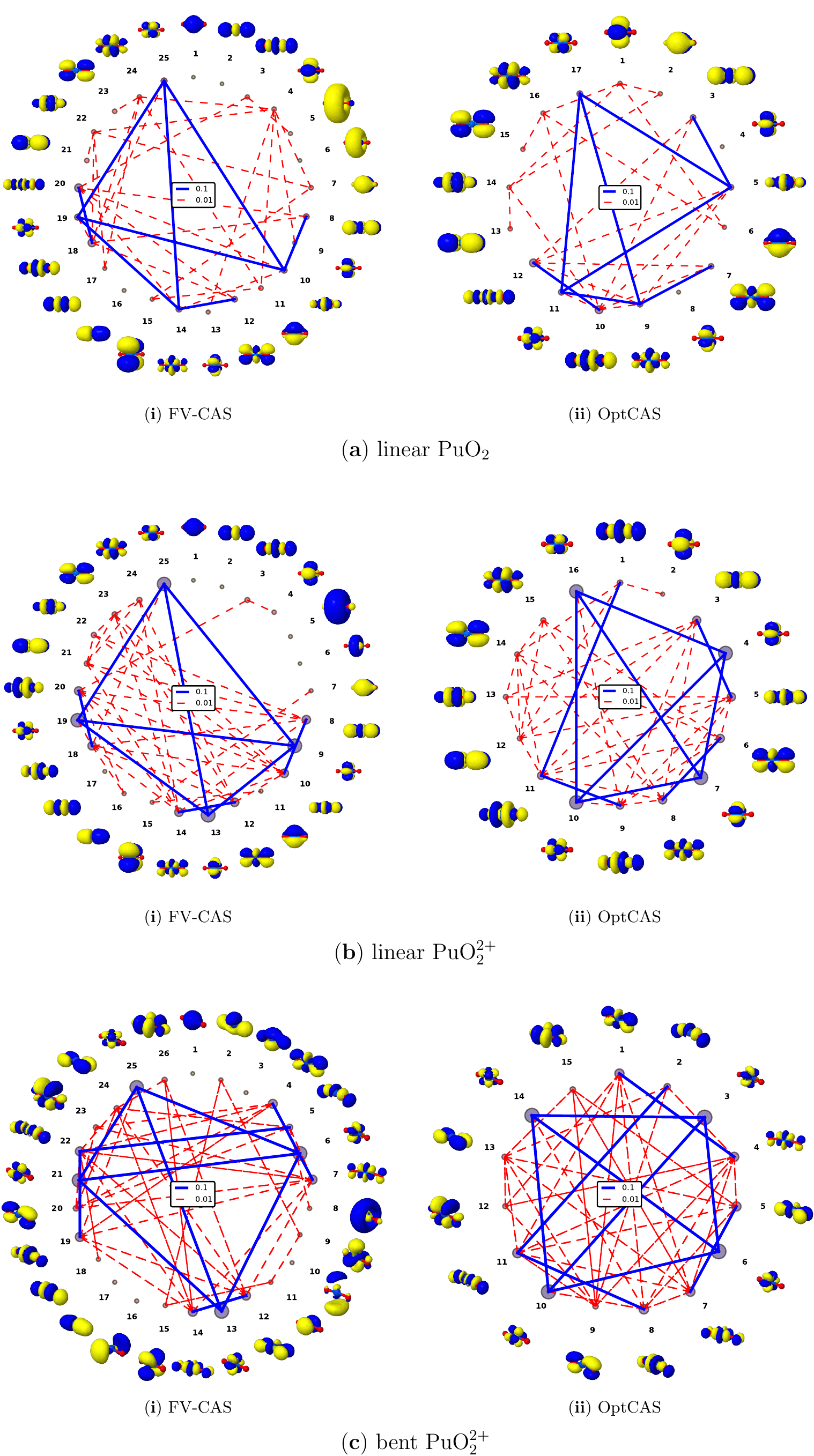}
\caption{Orbital correlations for the \ce{PuO2} molecule and linear and \textit{bent} \ce{PuO2^{2+}} ions and two different active spaces. The strength of the orbital-pair correlations is color-coded: the strongest orbital-pair correlations are marked by blue lines ($10^{-1}$), followed by orbital-pair correlations linked by red lines ($10^{-2}$).
} 
\label{fig:i12-1}
\end{figure*}
\subsection{Entanglement and correlation measures}
In order to select the most important active space orbitals for static/non-dynamic electron correlation, we used concepts of quantum information theory to quantify the entanglement and correlation of orbitals.
Specifically, the entanglement between one orbital and all remaining orbitals is quantified by the single-orbital entropy.
It can be calculated from the von Neumann entropy of the reduced density matrix of the orbital of interest, the so-called one-orbital reduced density matrix. In contrast to the $N$-particle reduced density matrix, which is defined for a constant number of particles, the one-orbital reduced density matrix is defined for a varying number of particles and hence its dimension is equal to the dimension of the one-orbital Fock space.
In the case of spatial orbitals, we have four different one-particle states (either unoccupied, singly occupied with a spin-up or spin-down electron, or doubly occupied orbitals) and the one-orbital reduced density matrix is a $4\times4$ matrix.
Its elements can be calculated from the one- and two-particle reduced density matrices~\cite{Kasia_ijqc} or from generalized correlation functions.~\cite{Barcza2013,entanglement_bonding_2013} We refer the interested reader to refs~\citenum{Ors_ijqc,entanglement_bonding_2013,barcza2014entanglement,Kasia_ijqc} for more details on how to calculate orbital-reduced density matrices.

The single-orbital entropy for orbital $i$ is determined from the eigenvalues of the one-orbital reduced density matrix $\omega_{\alpha;i}$,~\cite{DEAS}
\begin{equation}\label{eq:s1}
s_i=-\sum_{\alpha=1}^{4}{\omega_{\alpha;i} \ln \omega_{\alpha;i}},
\end{equation}
and thus represents the entanglement entropy of orbital $i$.
Similarly, the entropy of two orbitals within the orbital bath is quantified by the two-orbital reduced density matrix as
\begin{equation}
s_{i,j}=-\sum_{\alpha=1}^{16}{\omega_{\alpha;i,j}\ln \omega_{\alpha;i,j}},
\end{equation}
where $\omega_{\alpha;i,j}$ are the eigenvalues of the two-orbital reduced density matrix.
In contrast to the one-orbital reduced density matrix, the two-orbital reduced density matrix is defined in terms of basis states of a two-orbital Fock space, which contains 16 possible states in the case of spatial orbitals ($|$\zero \,\zero$\rangle$, $|$\zero \,\down$\rangle$,  $|$\down \,\zero$\rangle$, $|$\zero \,\up$\rangle$, ~\dots, $|$\double \,\double$\rangle$).

Given $s_i$ and $s_{i,j}$, we can quantify the correlation between two orbitals $i$ and $j$ by the orbital-pair mutual information,~\cite{Rissler2006,DEAS,Legeza2006,barcza2014entanglement}
\begin{equation}
 I_{i|j}=s_i+s_j-s_{i,j},
\end{equation}
which describes both quantum and classical correlations between two orbitals $i,j$.
In the following, we will use diagrams to represent $I_{i|j}$. Specifically, the strength of the orbital-pair mutual information is color-coded. Strongly correlated orbital pairs are connected by blue lines ($I_{i|j} \approx 10^{-1}$), while moderately correlated orbitals are linked by red lines ($I_{i|j} \approx 10^{-2}$), etc.
Moreover, we will seek active orbital spaces that allow for an accurate description of static/nondynamic electron correlation effects in all investigated plutonium oxides. One possibility to define stable and reliable active spaces in correlation calculations was proposed by some of us~\cite{entanglement_letter,Kasia_ijqc} and recently applied to facilitate black-box DMRG calculations.~\cite{Stein2016} In contrast to ref.~\citenum{Stein2016}, which uses the single-orbital entropy as exclusive selection criterion, we will exploit the orbital-pair mutual information in defining an optimal active space, primarily because $I_{i|j}$ allows us to quantify the correlation between orbital pairs and thus represents an immediate measure for electron correlation effects between orbital pairs embedded in an active space.
Since we are only interested in reproducing the largest orbital-pair correlations $I_{i|j}>10^{-2}$ that are important for nondynamic/static electron correlation, the orbital-pair correlation-based active space will be constructed by excluding all orbitals for which all values of the orbital-pair mutual information are smaller than $10^{-2}$.
If the resulting active orbital spaces are stable, we should recover the largest orbital-pair correlations with respect to the reference calculations.
Otherwise, the acceptance threshold for $I_{i|j}$ has to be further reduced (for instance, to $10^{-3}$).
In this work, a cutoff threshold for $I_{i|j}$ of $0.01$ was, however, sufficient in reproducing the dominant orbital-pair correlations and we did not perform additional calculations with smaller thresholds for $I_{i|j}$.
The error introduced by decreasing the size of the active space can be quantified by determining, for instance, the squared deviation of orbital-pair correlations for the optimized active orbital space, here optCAS, with respect to the reference active space, here FV-CAS,
\begin{equation}
 \varepsilon_{\mathrm{optCAS}} = \sum_{i,j \in \mathrm{optCAS}} \left(I_{i|j}^{\mathrm{FV-CAS}}-I_{i|j}^{\mathrm{optCAS}} \right)^2.
\end{equation}
In this work however, we focus only on the distribution of $I_{i|j}$ determined for different CASs to assess the reliability of the reduced active orbital spaces.
Finally, we should emphasize that we use two-orbital correlation measures to study the electronic structures and bonding patterns in plutonium oxides. An extension to multi-orbital correlations has been presented recently in Ref.~\citenum{Szalay2016}.

\section{Results and Discussion}\label{sec:results}
Since the detailed bonding mechanisms of the larger plutonium oxides \ce{PuO3} and \ce{PuO2(OH)2} are unknown, we will compare the properties on the \textit{bent}  \ce{PuO2^{2+}} molecule as a possible subunit of the \ce{PuO3} and \ce{PuO2(OH)2} complexes. As a preliminary step, we will focus on the linear \ce{PuO2} and \ce{PuO2^{2+}} species.
Specifically, we will scrutinize how the addition of the distant oxygen atom to the \ce{PuO2^{2+}} moiety influences orbital correlations and bonding patterns when going from \ce{PuO2^{2+}} to \ce{PuO3}.
Then, we will discuss how the electronic structure changes when two hydroxy groups are added to the \ce{PuO2^{2+}} subunit forming the \ce{PuO2(OH)2} molecule.


\subsection{The \ce{PuO2} molecule}

According to La Macchia~\textit{et al.}, spin-orbit coupling induces a change in the principle electronic configuration; at the spin-orbit level, it is 96~\% dominated by a $\mathrm{^5\phi_u}$ spin-free state, \textit{i.e.}, the dominant configuration is $\mathrm{7s^{(1)}}\delta_u^{(2)}\phi_u^{(2)}$.~\cite{ivan_puo2_08} However at the spin-free level, the ground state is a $\mathrm{^5\Sigma^+_g}$ that lies 1800~$\mathrm{cm^{-1}}$ below the $\mathrm{^5\phi_u}$ state, and corresponds to the single occupation of the $\delta_u$ and $\phi_u$ molecular orbital pairs. Thus, in this study we will consider only the spin-free $\mathrm{^5\Sigma^+_g}$ state and analyse its DMRG matrix product state wave function. The latter is described by a single determinant  as shown by the wave-function analysis at the CASSCF level, that could be decomposed into two major determinants, $\ket{\delta_u^{(1)}\delta_u^{(1)}\phi_u^{(1)}\phi_u^{(1)}}$  and $\ket{\phi_u^{(1)}\phi_u^{(1)}\pi_u^{(1)}\pi_u^{(1)}}$, with a weight of 90~\% and 8~\%, respectively.

The most important orbital-pair correlations for the linear \ce{PuO2} molecule and all investigated orbital spaces are shown in Fig.~\ref{fig:i12-1}(a).
The dominant orbital correlations as obtained from the optCAS calculation are in good agreement with the correlation diagram of the FV-CAS reference, indicating that static/nondynamic electron correlation effects are accurately captured in the optCAS active space.

\begin{figure}[!htbp]
\includegraphics[width=0.99\columnwidth]{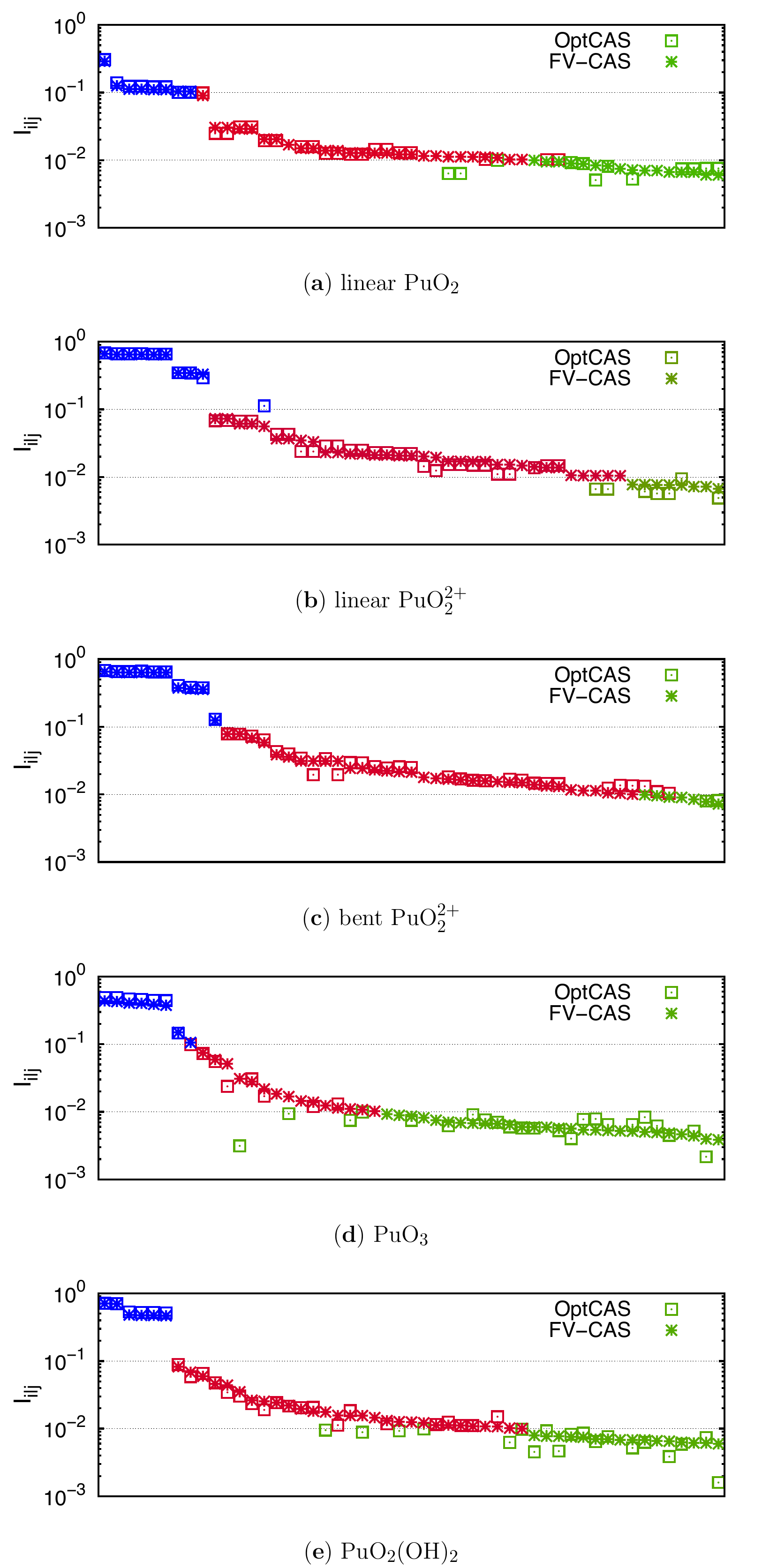}
\caption{Decaying values for the orbital-pair mutual information for all investigated plutonium oxide complexes and different active spaces. The values of the mutual information are ordered with respect to the FV-CAS reference calculation, \textit{that is}, each $I_{i|j}$ is plotted for the same indices $i$ and $j$ of FV-CAS and optCAS.}
\label{fig:i12decay}
\end{figure}

To highlight the good agreement in orbital correlations between FV-CAS and optCAS, Fig.~\ref{fig:i12decay} shows the sorted, decaying values of the orbital-pair mutual information for the first 50 strongly correlated orbital pairs.
For better comparison, we use the same color-coding scheme for the strength of the orbital-pair mutual information as displayed in Fig.~\ref{fig:i12-1}.
Each $I_{i|j}$ in Fig.~\ref{fig:i12decay} is shown for the same orbital pair $i,j$ and sorted with respect to the FV-CAS reference values.
As shown in Fig.~\ref{fig:i12decay}(a), optCAS results in a distribution of orbital correlations that is similar to the FV-CAS reference distribution, with 8 dominant distributions over $\mathrm{10^{-1}}$, and one at the limit with a squared deviation of $\varepsilon_{\mathrm{optCAS}}=0.00128$.

For both optCAS and FV-CAS, the seven of the eight distributions correspond to strong pair-correlation between the bonding  and antibonding $\pi_u/\pi_u^*$  orbitals (nos.~3, 5, 7 and 9 in Fig.~\ref{fig:i12-1}(a-ii)) (2 blue lines) as well as the antibonding $\pi_u/\pi_u^*$  orbitals  and the two nonbonding ${\delta}_u$ orbitals (nos.~11 and 17 in Fig.~\ref{fig:i12-1}(a-ii); 4 blue lines) and the two nonbonding ${\delta}_u$ (1 blue line). Specifically, the bonding and antibonding combination of both oxygen $p_z$ orbitals and the plutonium $f_{z^3}$ orbital lead to the formation of a $\sigma_u$ bond between the plutonium center and the oxygen atoms, which results in one strongly correlated $\sigma_u$--$\sigma_u^*$ orbital pair (nos.~10 and 12 in Fig.~\ref{fig:i12-1}(a-ii)).
Note that the doubly degenerate ${\phi}_u$ orbitals have $I_{i|j}$ values much smaller than $10^{-3}$. Even so, the entanglement with other antibonding orbitals remains weak, the population analysis (cf Table S2 in the ESI) exhibits a single occupation of the ${\phi}_u$ orbitals, making them obviously important in the description of the ground-state. 
Thus one can conclude that the remaining $\sigma_u/\sigma_u^*$, $\pi_u/\pi_u^*$, and ${\delta}_u$ as well as ${\phi}_u$ orbitals are important to describe nondynamic/static electron correlation appropriately. The last statement conforms to the choice of La Macchia \textit{et al.}~not to consider the $\pi_g$ and $\sigma^*_g$ molecular orbitals in their active space.~\cite{ivan_puo2_08}\\


\subsection{The plutonyl \ce{PuO2^{2+}} ion}

\subsubsection{The linear structure} 
The electronic structure of the linear \ce{PuO2^{2+}} complex was optimized in its triplet ground state that consists of a linear combination of two determinants with equal weights, $\ket{\phi_u^{(1)}\mathrm{(B_{3u})}\delta_u^{(1)}\mathrm{(B_{1u})}}$ and $\ket{\phi_u^{(1)}\mathrm{(B_{2u})}\delta_u^{(1)}\mathrm{(A_{u})}}$.
The orbital-pair correlations are displayed in Fig.~\ref{fig:i12-1}(b) for both active spaces studied.
In the optCAS calculation, the orbital-pair correlations are identical to that of the large FV-CAS reference calculation (compare Figs.~\ref{fig:i12-1}(b-i) and (b-ii)).
The good agreement of optCAS and FV-CAS in describing static/nondynamic electron correlation is also evident in the decay of the orbital-pair mutual information shown in Fig.~\ref{fig:i12decay}(b).
In general, differences in orbital-pair correlations between optCAS and FV-CAS are negligible.
The largest differences are found for weakly correlated orbital-pairs, which can be attributed to dynamic electron correlation effects beyond the optCAS orbitals, and between the ${\sigma}_g$ and ${\sigma}_g^*$ orbitals (nos.~1 and 11 in Fig.~\ref{fig:i12-1}(b-ii)), which are slightly over-correlated in the optCAS calculation as compared to the FV-CAS and lead to a larger squared deviation of $\varepsilon_{\mathrm{optCAS}}=0.00800$ compared to \ce{PuO2}.
Nonetheless, similar $I_{i|j}$ profiles for optCAS and FV-CAS suggest the proper choice of the optCAS active space.

Similarly to the linear \ce{PuO2} molecule, the strongly-correlated orbital pairs are the singly-occupied ${\delta}_u$ (nos.~10 and 16 in Fig.~\ref{fig:i12-1}(b-ii)) orbitals as well as the bonding and antibonding combination of the ${\pi}_u$ orbitals (nos.~3 and 5 as well as 6 and 8 in Fig.~\ref{fig:i12-1}(b-ii)) and the ${\sigma}_u$ orbitals (nos.~9 and 11 in Fig.~\ref{fig:i12-1}(b-ii)). However, in contrast to \ce{PuO2}, the doubly-degenerate (and nonbonding) ${\phi}_u$ orbitals (nos.~4 and 7 in Fig.~\ref{fig:i12-1}(b-ii)) become strongly correlated with the ${\delta}_u$ orbitals, while the pair-correlation between the ${\pi}_u^*$ and ${\delta}_u$ decreases with $I_{i|j}$ values below $\mathrm{10^{-2}}$.
Furthermore, like in linear \ce{PuO2}, the ${\pi}_u$ and ${\pi}_u^*$ orbitals (nos.~3 and 5 as well as 6 and 8 in Fig.~\ref{fig:i12-1}(b-ii)) are more strongly correlated than the ${\pi}_g$--${\pi}_g^*$ orbital pair (nos.~12 and 13 as well as 14 and 15 in Fig.~\ref{fig:i12-1}(b-ii)).
The different strength of correlation between the ${\pi}_u$/${\pi}_u^*$ and ${\pi}_g$/${\pi}_g^*$ orbitals is also observable in the abrupt decrease in orbital-pair mutual information around $10^{-1}$ (see Fig.~\ref{fig:i12decay}(b)).
Finally, to account properly for static/nondynamic electron correlation, a CASSCF calculation should include, as for the \ce{PuO2} molecule, the $\sigma_u/\sigma_u^*$, $\pi_u/\pi_u^*$, and ${\delta}_u$ as well as ${\phi}_u$ orbitals.

\subsubsection{The bent structure} 
For simplicity and in the following sections, the orbital labels were kept identical to those referring to $\mathrm{D_{2h}}$ symmetry.	
Distorting the structure of the linear \ce{PuO2^{2+}} molecule into the \textit{bent} \ce{PuO2^{2+}} subunit present in \ce{PuO3} and \ce{PuO2(OH)2} changes the orbital-pair correlation picture marginally. As observed for the linear \ce{PuO2}/\ce{PuO2^{2+}}, the strongly correlated orbital pairs are the singly-occupied ${\delta}_u$ and ${\phi}_u$ (nos.~3, 6, 10, and 14 in Fig.~\ref{fig:i12-1}(c-ii)) orbitals as well as both $\pi_u/{\pi_u^*}$ orbitals (nos.~1 and 4 as well as 5 and 7 in Fig.~\ref{fig:i12-1}(c-ii)) and the $\sigma_u/{\sigma_u^*}$ orbitals (nos.~8 and 11 in Fig.~\ref{fig:i12-1}(c-ii)).
Furthermore, the ${\pi}_g$--${\pi}_g^*$ orbital pairs (nos.~9 and 12 as well as 13 and 15 in Fig.~\ref{fig:i12-1}(c-ii)) remain moderately correlated as in the linear \ce{PuO2^{2+}} counterpart.
Bending the \ce{Pu\bond{#}O} bond increases, however, the orbital-pair correlation between the ${\sigma}_g$ and ${\sigma}_u^*$ orbitals (nos.~2 and 11 in Fig.~\ref{fig:i12-1}(c-ii)) by approximately one order of magnitude.
Thus, in order to describe the strongest orbital-pair correlations for \textit{bent} \ce{PuO2^{2+}}, the ${\sigma}_g$ orbital has to be included in the active space.
As observed above, the differences in orbital-pair correlations between optCAS and FV-CAS are negligible with $\varepsilon_{\mathrm{optCAS}}=0.00523$.

\begin{figure*}[!htbp]
\centering
\centering
\includegraphics[width=0.99\textwidth]{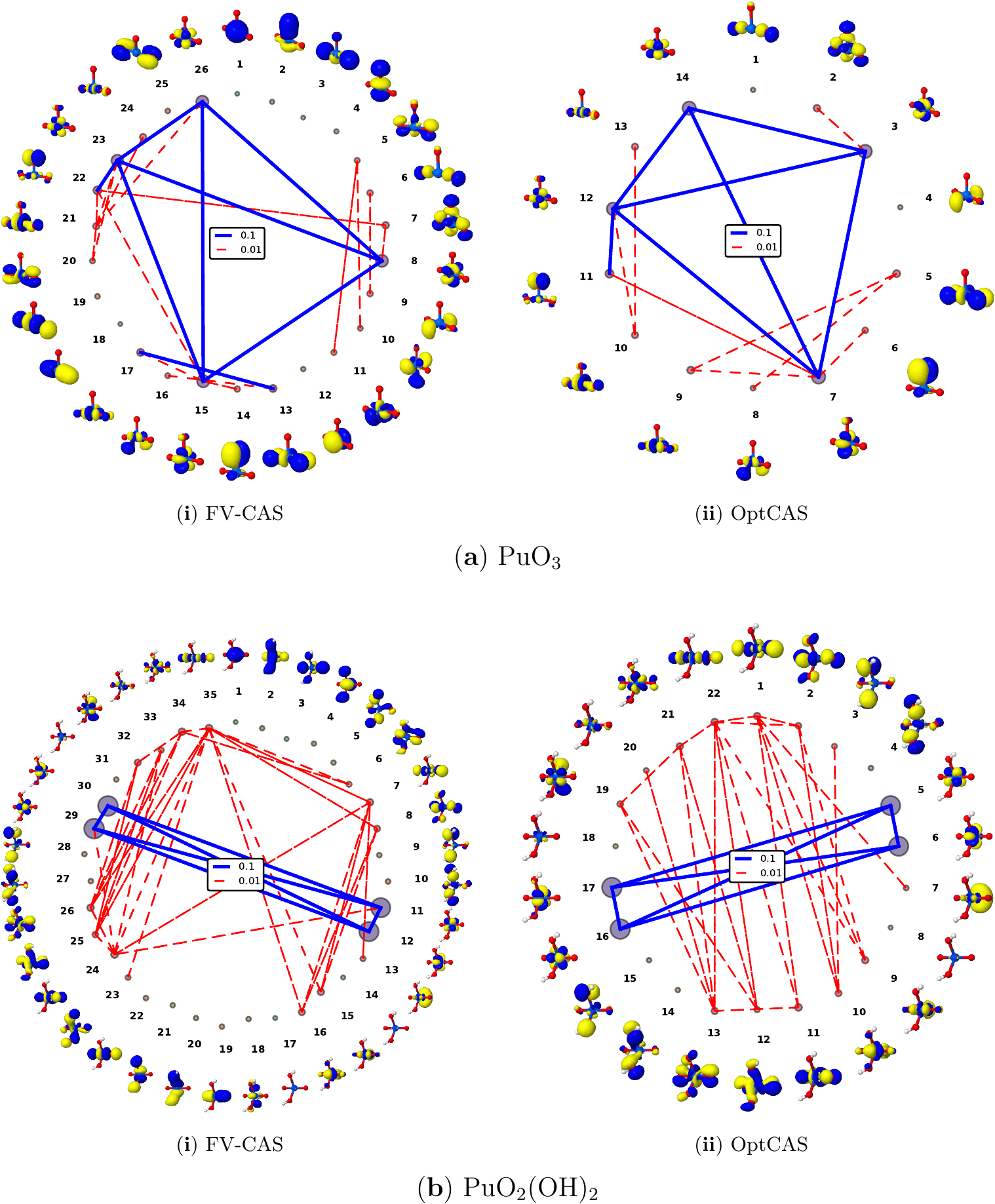}
\caption{Orbital correlations for the \ce{PuO3} and \ce{PuO2(OH)2} molecules and two different active spaces. The strength of the orbital-pair correlations is color-coded: the strongest orbital-pair correlations are marked by blue lines ($10^{-1}$), followed by orbital-pair correlations linked by red lines ($10^{-2}$).
} 
\label{fig:i12-2}
\end{figure*}

The most pronounced differences in orbital-pair correlations can be found for $\sigma$-type orbitals.
Since, however, only one orbital-pair correlation is affected (cf., Figs.~\ref{fig:i12-2}(b) and (c)), the structural distortion can be considered too small to have a significant impact on the $\sigma$- and $\pi$-bonding mechanisms in \ce{PuO2^{2+}}.

\subsection{The \ce{PuO3} molecule}
In contrast to structural distortions, addition of the distant O$^{2-}$ group to the  \textit{bent} \ce{PuO2^{2+}} subunit changes the orbital-pair correlations.
Similar to the \textit{bent} \ce{PuO2^{2+}} subunit, the plutonium ${\delta}_u$ and ${\phi}_u$ orbitals of \ce{PuO3} are highly correlated with each other, while the remaining active space orbitals are moderately to weakly correlated (see Fig.~\ref{fig:i12-2}(a)), especially the $\sigma_u$--$\sigma_u^*$ and $\pi_u$--$\pi_u^*$ orbital pairs that decrease below the threshold of $\mathrm{10^{-1}}$ in the optCAS.  Although weak correlations ($I_{i|j}\le 10^{-2}$) are underestimated in our optCAS calculations compared to the FV-CAS reference (see Fig.~\ref{fig:i12decay}(c)), optCAS still represents the smallest active space that contains all important orbitals to reliably describe nondynamic/static electron correlation.
We should note, however, that the squared deviation increases to $\varepsilon_{\mathrm{optCAS}}=0.02389$ because we have increased the cutoff threshold for the orbital-pair mutual information to $I_{i|j}\approx0.02$, excluding orbital no.~5 in Fig.~\ref{fig:i12-2}(a-ii).

The orbital-pair correlation diagrams can be further used to elucidate how the bonding mechanism in \ce{PuO3} changes compared to the \ce{PuO2^{2+}} subunit.
In the \ce{PuO3} molecule, the singly occupied plutonium ${\delta}_u$ and ${\phi}_u$ orbitals are strongly correlated (nos.~3, 7, 12, and 14 in Fig.~\ref{fig:i12-2}(a-ii)), with orbital-pair mutual information similar in magnitude to the ${\delta}_u$/${\phi}_u$ correlations in \ce{PuO2^{2+}}.
Furthermore, in contrast to the strongly correlated $\sigma_u-\sigma_u^*$ orbital pair in the \ce{PuO2}/\ce{PuO2^{2+}} molecules, the $\sigma_u-\sigma_u^*$ orbitals are only moderately correlated (nos.~10 and 13 in Fig.~\ref{fig:i12-2}(a-ii)) in \ce{PuO3}.

We observe that only the ${\pi}_u$ and ${\pi}_u^*$ orbitals out of the plane defined by the molecule are still strongly correlated (nos.~13 and 17 in Fig.~\ref{fig:i12-2}(a-i) with 1 blue line and nos.~5 and 9 in Fig.~\ref{fig:i12-2}(a-ii) with 1 red line), while the ${\pi}_u$ and ${\pi}_u^*$ orbitals located in the plane  are only weakly correlated (nos.~5 and 20 in Fig.~\ref{fig:i12-2}(a-i), not shown in the Figure). More striking is the important orbital-pair correlation between the $p_z$ orbital of the distant oxygen with the $\delta_u$ orbital (nos.~22 and 23 in Fig.~\ref{fig:i12-2}(a-i)), and the weaker ones with the planar bonding ${\pi}_u$ orbital (no.~20 in Fig.~\ref{fig:i12-2}(a-i)) and the hybrid antibonding orbital constructed as a combination of the planar $\pi_u^*$ MO and the distant-oxygen $p_x$ orbital (no.~2 in Fig.~\ref{fig:i12-2}(a-ii)); the later  being coupled  with the $\phi_u$ orbital (no.~3 in Fig.~\ref{fig:i12-2}(a-i))

An interesting feature concerns the orbitals that describe the bonding between the distorted plutonyl subunits and the distant oxygen. The bonding and antibonding orbitals forming $\sigma$-type orbitals (nos.~2 and 4 in Fig.~\ref{fig:i12-2}(a-i)) do not contribute at all in the description of the ground-state wave-function.
Hence, we can assume that the axial $\pi$ bond is not formed by the plutonium ${\pi}_u$ and oxygen $p_z$ orbitals.
Instead, the axial and equatorial oxygen $p_z$ orbitals interact with the plutonium $6d_{\sigma}$ orbital (no.~2 in Fig.~\ref{fig:i12-2}(a-ii)) and the corresponding molecular orbital can be considered as a hybrid of an axial plutonium--oxygen $\sigma$ bond and ${\delta}_u$-type bonding between the three oxygen atoms.
This $\sigma$/${\delta}_u$-type molecular orbital is moderately correlated with the singly occupied plutonium ${\phi}_u$ orbital.
The remaining $p$ orbitals of the axial oxygen atom slightly mix with plutonium 6d orbitals (nos.~6 and 11 in Fig.~\ref{fig:i12-2}(a-ii)) and become moderately correlated with the plutonium ${\phi}_u$ and ${\delta}_u$ orbitals.

The strong discrepancies in $I_{i|j}$ between optCAS and FV-CAS can be explained by investigating the electronic structure of \ce{PuO3}.
In contrast to the \ce{PuO2^{2+}} subunit, the electronic structure of \ce{PuO3} is dominated by dynamic electron correlation effects beyond the optCAS orbitals, which results in $I_{i|j}$ being underestimated in comparison to the FV-CAS reference distribution (see Fig.~\ref{fig:i12decay}(c)).
However, the missing dynamic electron correlation effects do not considerably affect the nondynamic/static correlation in the (optCAS) active space and can be included \textit{a posteriori} using, for instance, perturbation theory.


Note that nondynamic/static electron correlation effects can be accurately described within the optCAS, which contains the distant-oxygen $p$ orbitals and the other orbitals of the distorted plutonyl subunit. Hence, a minimal active space containing the $\sigma_u/\sigma_u^*$, $\pi_u/\pi_u^*$, and ${\delta}_u$ as well as ${\phi}_u$ orbitals and the $p$-shell of the oxygen would suffice to accurately model the ground-state wave function in \ce{PuO3}.

\subsection{The \ce{PuO2(OH)2} complex}
As observed for the \ce{PuO3} complex, extending the ligand sphere of \ce{PuO2^{2+}} with two hydroxy groups changes the orbital-pair correlations compared to the triatomic subunit.
In the \ce{PuO2(OH)2} molecule, the singly-occupied ${\delta}_u$ and ${\phi}_u$ orbitals (nos.~5, 6, 16, and 17 in Fig.~\ref{fig:i12-2}(b-ii)) are strongly correlated, while the plutonium ${\pi}_g$ and ${\pi}_u$ orbitals are moderately correlated.
To describe nondynamic/static electron correlation appropriately, the active space can be reduced from 35 (FV-CAS) to 22 (OptCAS) orbitals without changing orbital-pair correlations (see also Fig.~\ref{fig:i12decay}(d)).
Although some orbital-pair correlations with $I_{i|j}\approx10^{-2}$ are underestimated, the orbital-pair mutual information for $I_{i|j}>10^{-2}$ determined from optCAS agrees well with the FV-CAS reference distribution (see Fig.~\ref{fig:i12decay}(d)) with a squared deviation of $\varepsilon_{\mathrm{optCAS}}=0.01030$.
The discrepancies in $I_{i|j}$ between optCAS and FV-CAS can be attributed to dynamic correlations effects outside the optCAS active space.
However, they do not influence the orbital-pair correlations where $I_{i|j}\gg10^{-2}$.
Thus, optCAS with 20 electrons correlated in 22 orbitals represents a good choice to describe the most important correlation effects present in the \ce{PuO2(OH)2} molecule.
Note that optCAS could be further reduced by excluding 5 molecular orbitals (nos.~4, 8, 14, 15, and 18 in Fig.~\ref{fig:i12-2}(b-ii)) as those orbitals are not important for static/nondynamic electron correlation. Since, however, those orbitals lie energetically between the other active space orbitals, we have kept them in our optCAS calculations.
If we neglect changes in orbital-pair correlations due to minor structural differences in the \ce{PuO2^{2+}} subunit, we can conclude that the addition of two hydroxy ligands decreases nondynamic/static electron correlation effects in the \ce{PuO2(OH)2} complex compared to the bare \ce{PuO2^{2+}} model compound (compare Figs.~\ref{fig:i12decay}(b) and (d)).
Furthermore, we observe the most profound jump in the mutual information around $I_{i|j}\approx10^{-2}$ in the \ce{PuO2(OH)2} molecule which stresses how additional ligands influence orbital-pair correlations in the \ce{PuO2^{2+}} unit.

In contrast to the \ce{PuO2} and \ce{PuO2^{2+}} units, the plutonium 5f and 6d orbitals hybridize considerably with each other as well as with all oxygen $p$ orbitals.
Specifically, the plutonium 5f/6d orbitals mix with the oxo $p$ orbitals to form moderately correlated $f/d_\pi$ and $f/d_{\pi^*}$-type orbitals (nos.~9, 13, 21, and 22 in Fig.~\ref{fig:i12-2}(b-ii)), while the hydroxy $p$ orbitals mix with the oxo $p$ orbitals and the plutonium 5f/6d orbitals to form molecular orbitals (nos.~2 and 12 in Fig.~\ref{fig:i12-2}(b-ii)) that are moderately correlated with the $f/d_\pi$ and $f/d_{\pi^*}$-type orbitals.
This reduces the ${\pi}_u$--${\pi}_u^*$ orbital-pair correlations as observed in the \ce{PuO2^{2+}} unit.
Furthermore, the orbital-pair correlation of the $\sigma$--$\sigma^*$ orbital pair (nos.~1 and 11 in Fig.~\ref{fig:i12-2}(b-ii)) weakens when going from the bare \ce{PuO2^{2+}} molecule to \ce{PuO2(OH)2}.
Finally, the hydroxy lone-pairs (nos. 3, 4, 14, and 15 in Fig.~\ref{fig:i12-2}(b-ii)) are only weakly correlated ($I_{i|j}<10^{-2}$) with molecular orbitals centered on the \ce{PuO2^{2+}} subunit, which are thus not shown in the Figure.

Therefore, we can conclude that the $\pi$ bonding mechanism changes most significantly when two hydroxy groups are added to the bare \ce{PuO2^{2+}} complex.
A similar observation was already made for the \ce{PuO3} compound discussed above.
Specifically, the plutonium 5f and 6d orbitals mix considerably and form bonding and antibonding $f/d_\pi$ orbitals with the oxo $p$ orbitals.
These hybrid $f/d_\pi$ orbitals contribute to the $\pi$ bonding in the \ce{PuO2^{2+}} subunit of the \ce{PuO2(OH)2} molecule.
Furthermore, the orbital-pair correlation diagrams suggest that the hydroxy ligands mainly interact with the \ce{PuO2^{2+}} subunit through hydroxo $p$ and plutonium $f/d$ mixing which alters the ${\pi}_g$ bonding interaction of the bare \ce{PuO2^{2+}} molecule and reduces the ${\pi}_g$--${\pi}_g^*$ orbital-pair correlation in the hydroxo-ligated \ce{PuO2^{2+}} complex.
Thus, the bonding mechanisms in the \ce{PuO2(OH)2} compound are a complex interplay of plutonium 5f and 6d orbitals, the oxo $2p$ orbitals, and the oxygen $2p$ orbitals of the hydroxo ligands.

\normalem
\section{Conclusions}\label{sec:conclusions}
Actinide-containing compounds are remarkably challenging for present-day quantum chemistry, primarily because of the large number of the energetically close 5f, 6d, and 7s orbitals that have to be included in active space calculations as well as the need to account for relativistic effects in the quantum-chemical model.
In this work, we have investigated the electronic structure and bonding mechanisms of different plutonium oxides using the DMRG algorithm and concepts of quantum information theory.

For each plutonium compound, we studied two different active orbital spaces: a large (full-valence) active space (FV-CAS) and an active space that reproduces the most important nondynamic/static orbital-pair correlations of this FV-CAS reference calculation, but simultaneously, allows us to reduce the number of active space orbitals as much as possible (here, called optCAS).
The orbital-pair mutual information, used to select the most important active space orbitals from a large active space reference calculation, allows us to dissect electron correlation effects and easily identify those orbitals that are important for nondynamic/static and weak correlation, respectively.~\cite{entanglement_letter}

Thus, an optimal active space can be defined by only selecting the strongly correlated orbitals,  from a large active space calculation that reproduces the orbital-pair correlation diagram of the reference calculation, resulting in DMRG(14,16) for linear \ce{PuO2^{2+}}, DMRG(14,15) for \textit{bent} \ce{PuO2^{2+}}, DMRG(18,17) for linear \ce{PuO2}, DMRG(14,14) for \ce{PuO3}, and finally DMRG(20,22) for \ce{PuO2(OH)2}.
Most importantly, the discrepancies in $I_{i|j}$ between optCAS and FV-CAS are minor and are caused by dynamic electron correlation effects beyond the optCAS active space orbitals.
Thus, optCAS should result in reliable zeroth-order wave function for an \textit{a posteriori} treatment of dynamic electron correlation effects.

Our results also serve to underscore the differences in the bonding picture for \ce{PuO3} and \ce{PuO2(OH)2} with respect to the bare plutonyl species.
For both molecules, the $\pi$-bonding mechanism changes significantly when oxo or hydroxo ligands are added to the triatomic plutonyl unit.
Specifically for \ce{PuO2(OH)2}, we observe considerable changes with respect to plutonyl, with substantial mixing of the plutonium 5f and 6d orbitals that results in two bonding and anti bonding hybrid $f/d_\pi$ orbitals, with the orbital-pair correlation analysis indicating that the hydroxo and plutonyl subunits interact through mixing of plutonium $f/d$ and hydroxo $p$ orbitals, which
reduces the ${\pi}_g$--${\pi}_g^*$ correlation found in \ce{PuO2^{2+}}.
Thus, concepts of quantum information theory represent a rich and useful tool to perform electronic structure calculations, to interpret electronic wave functions, and to gain chemical insights into the structure of molecules that are difficult to understand using standard approaches like molecular orbital diagrams.
Most importantly, our approach can be reliably applied whenever the electronic wave function can be accurately optimized, even in cases when the simple picture of interacting orbitals completely fails.

\section{Acknowledgement}
We gratefully acknowledge financial support from the France-Canada Research Fund, Natural Sciences and Engineering Research Council of Canada (NSERC), 
and the Hungarian Research Fund (OTKA K100908 and NN110360). The members of the PhLAM laboratory acknowledge support from the CaPPA project (Chemical and Physical Properties of the Atmosphere) that is funded by the French National Research Agency (ANR) through the PIA (Programme d'Investissement d'Avenir) under contract "ANR-11-LABX-0005-01" and by the Regional Council ``Hauts de France'' and the  "European Funds for Regional Economic Development (FEDER)
C.D.~acknowledges financial support from the NSERC Undergraduate Student Research Award fellowship.
K.B.~acknowledges financial support from a SONATA BIS grant of the National Science Centre, Poland (no.~2015/18/E/ST4/00584) and a Marie-Sk\l{}odowska-Curie Invididual Fellowship project no.~702635--PCCDX. 
P.T.~thanks the National Science Center Grant No.~DEC-2013/11/B/ST4/00771 and No.~DEC-2012/07/B/ST4/01347. 

The authors acknowledge support for computational resources from \textsc{SHARCNET}, a partner consortium in the Compute Canada national HPC platform.
Calculations have been carried out using resources provided by Wroclaw Centre for Networking and Supercomputing (http://wcss.pl), grant No.~10105802.





\bibliography{rsc} 

\providecommand*{\mcitethebibliography}{\thebibliography}
\csname @ifundefined\endcsname{endmcitethebibliography}
{\let\endmcitethebibliography\endthebibliography}{}
\begin{mcitethebibliography}{80}
\providecommand*{\natexlab}[1]{#1}
\providecommand*{\mciteSetBstSublistMode}[1]{}
\providecommand*{\mciteSetBstMaxWidthForm}[2]{}
\providecommand*{\mciteBstWouldAddEndPuncttrue}
  {\def\EndOfBibitem{\unskip.}}
\providecommand*{\mciteBstWouldAddEndPunctfalse}
  {\let\EndOfBibitem\relax}
\providecommand*{\mciteSetBstMidEndSepPunct}[3]{}
\providecommand*{\mciteSetBstSublistLabelBeginEnd}[3]{}
\providecommand*{\EndOfBibitem}{}
\mciteSetBstSublistMode{f}
\mciteSetBstMaxWidthForm{subitem}
{(\emph{\alph{mcitesubitemcount}})}
\mciteSetBstSublistLabelBeginEnd{\mcitemaxwidthsubitemform\space}
{\relax}{\relax}

\bibitem[Hartmann(2012)]{Hartmann2012}
J.~Hartmann, \emph{Nat. Chem.}, 2012, \textbf{4}, 1052\relax
\mciteBstWouldAddEndPuncttrue
\mciteSetBstMidEndSepPunct{\mcitedefaultmidpunct}
{\mcitedefaultendpunct}{\mcitedefaultseppunct}\relax
\EndOfBibitem
\bibitem[Haire and Haschke(2001)]{Haire2001}
R.~G. Haire and J.~M. Haschke, \emph{MRS Bull.}, 2001, \textbf{26},
  689--696\relax
\mciteBstWouldAddEndPuncttrue
\mciteSetBstMidEndSepPunct{\mcitedefaultmidpunct}
{\mcitedefaultendpunct}{\mcitedefaultseppunct}\relax
\EndOfBibitem
\bibitem[Nash(1993)]{PUREX1}
K.~L. Nash, \emph{Solvent~Extr.~Ion~ Exch.}, 1993, \textbf{11}, 729--768\relax
\mciteBstWouldAddEndPuncttrue
\mciteSetBstMidEndSepPunct{\mcitedefaultmidpunct}
{\mcitedefaultendpunct}{\mcitedefaultseppunct}\relax
\EndOfBibitem
\bibitem[Nash \emph{et~al.}(2000)Nash, Barrans, Chiarizia, Dietz, Jensen,
  Rickert, Moyer, Bonnesen, Bryan, and Sachleben]{PUREX2}
K.~L. Nash, R.~E. Barrans, R.~Chiarizia, M.~L. Dietz, M.~Jensen, P.~Rickert,
  B.~A. Moyer, P.~V. Bonnesen, J.~C. Bryan and R.~A. Sachleben,
  \emph{Solvent~Extr.~Ion~ Exch.}, 2000, \textbf{18}, 605--631\relax
\mciteBstWouldAddEndPuncttrue
\mciteSetBstMidEndSepPunct{\mcitedefaultmidpunct}
{\mcitedefaultendpunct}{\mcitedefaultseppunct}\relax
\EndOfBibitem
\bibitem[Ronchi \emph{et~al.}(2000)Ronchi, Capone, Colle, and
  Hiernaut]{Ronchi-JNM2000}
C.~Ronchi, F.~Capone, J.~Y. Colle and J.~P. Hiernaut, \emph{J. Nucl. Mat.},
  2000, \textbf{280}, 111--115\relax
\mciteBstWouldAddEndPuncttrue
\mciteSetBstMidEndSepPunct{\mcitedefaultmidpunct}
{\mcitedefaultendpunct}{\mcitedefaultseppunct}\relax
\EndOfBibitem
\bibitem[Reiher and Wolf(2009)]{Reiher_book}
M.~Reiher and A.~Wolf, \emph{Relativistic Quantum Chemistry. {T}he Fundamental
  Theory of Molecular Science}, Wiley, Dordrecht, 2009\relax
\mciteBstWouldAddEndPuncttrue
\mciteSetBstMidEndSepPunct{\mcitedefaultmidpunct}
{\mcitedefaultendpunct}{\mcitedefaultseppunct}\relax
\EndOfBibitem
\bibitem[Tecmer \emph{et~al.}(2016)Tecmer, Boguslawski, and
  K{\c{e}}dziera]{Tecmer2016}
P.~Tecmer, K.~Boguslawski and D.~K{\c{e}}dziera, in \emph{Handbook of
  Computational Chemistry}, ed. J.~Leszczy{\'{n}}ski, Springer Netherlands,
  Dordrecht, 2016, pp. 1--43\relax
\mciteBstWouldAddEndPuncttrue
\mciteSetBstMidEndSepPunct{\mcitedefaultmidpunct}
{\mcitedefaultendpunct}{\mcitedefaultseppunct}\relax
\EndOfBibitem
\bibitem[Vallet \emph{et~al.}(2000)Vallet, Maron, Teichteil, and
  Flament]{EPCISO}
V.~Vallet, L.~Maron, C.~Teichteil and J.-P. Flament, \emph{J.~Chem.~Phys.},
  2000, \textbf{113}, 1391--1402\relax
\mciteBstWouldAddEndPuncttrue
\mciteSetBstMidEndSepPunct{\mcitedefaultmidpunct}
{\mcitedefaultendpunct}{\mcitedefaultseppunct}\relax
\EndOfBibitem
\bibitem[Schreckenbach and Shamov(2010)]{schreckenbach_dft}
G.~Schreckenbach and G.~A. Shamov, \emph{Acc.~Chem.~Res.}, 2010, \textbf{43},
  19--29\relax
\mciteBstWouldAddEndPuncttrue
\mciteSetBstMidEndSepPunct{\mcitedefaultmidpunct}
{\mcitedefaultendpunct}{\mcitedefaultseppunct}\relax
\EndOfBibitem
\bibitem[Gomes \emph{et~al.}(2015)Gomes, R\'{e}al, Schimmelpfennig, Wahlgren,
  and Vallet]{gomes2015applied}
A.~S.~P. Gomes, F.~R\'{e}al, B.~Schimmelpfennig, U.~Wahlgren and V.~Vallet, in
  \emph{Applied Computational Actinide Chemistry}, ed. M.~Dolg, John Wiley \&
  Sons Ltd, Chichester, 2015, ch.~11, pp. 269--298\relax
\mciteBstWouldAddEndPuncttrue
\mciteSetBstMidEndSepPunct{\mcitedefaultmidpunct}
{\mcitedefaultendpunct}{\mcitedefaultseppunct}\relax
\EndOfBibitem
\bibitem[Maron \emph{et~al.}(1999)Maron, Leininger, Schimmelpfennig, Vallet,
  Heully, Teichteil, Gropen, and Wahlgren]{Maron-pu022+-1999}
L.~Maron, T.~Leininger, B.~Schimmelpfennig, V.~Vallet, J.-L. Heully,
  C.~Teichteil, O.~Gropen and U.~Wahlgren, \emph{Chem. Phys.}, 1999,
  \textbf{244}, 195--201\relax
\mciteBstWouldAddEndPuncttrue
\mciteSetBstMidEndSepPunct{\mcitedefaultmidpunct}
{\mcitedefaultendpunct}{\mcitedefaultseppunct}\relax
\EndOfBibitem
\bibitem[Clavagu\'{e}ra-Sarrio \emph{et~al.}(2004)Clavagu\'{e}ra-Sarrio,
  Vallet, Maynau, and Marsden]{Clavaguera-Sarrio_2004}
C.~Clavagu\'{e}ra-Sarrio, V.~Vallet, D.~Maynau and C.~J. Marsden,
  \emph{J.~Chem.~Phys.}, 2004, \textbf{121}, 5312--5321\relax
\mciteBstWouldAddEndPuncttrue
\mciteSetBstMidEndSepPunct{\mcitedefaultmidpunct}
{\mcitedefaultendpunct}{\mcitedefaultseppunct}\relax
\EndOfBibitem
\bibitem[Ismail \emph{et~al.}(1999)Ismail, Heully, Saue, Daudey, and
  Marsden]{Trond_1999}
N.~Ismail, J.-L. Heully, T.~Saue, J.-P. Daudey and C.~J. Marsden,
  \emph{Chem.~Phys.~Lett.}, 1999, \textbf{300}, 296\relax
\mciteBstWouldAddEndPuncttrue
\mciteSetBstMidEndSepPunct{\mcitedefaultmidpunct}
{\mcitedefaultendpunct}{\mcitedefaultseppunct}\relax
\EndOfBibitem
\bibitem[Hay \emph{et~al.}(2000)Hay, Martin, and Schreckenbach]{Hay2000}
P.~J. Hay, R.~L. Martin and G.~Schreckenbach, \emph{J. Phys. Chem. A}, 2000,
  \textbf{104}, 6259--6270\relax
\mciteBstWouldAddEndPuncttrue
\mciteSetBstMidEndSepPunct{\mcitedefaultmidpunct}
{\mcitedefaultendpunct}{\mcitedefaultseppunct}\relax
\EndOfBibitem
\bibitem[Odoh and Schreckenbach(2010)]{U_core_potentials}
S.~O. Odoh and G.~Schreckenbach, \emph{J.~Phys.~Chem.~A}, 2010, \textbf{114},
  1957\relax
\mciteBstWouldAddEndPuncttrue
\mciteSetBstMidEndSepPunct{\mcitedefaultmidpunct}
{\mcitedefaultendpunct}{\mcitedefaultseppunct}\relax
\EndOfBibitem
\bibitem[Kov{\'a}cs and Konings(2011)]{Kovacs_11}
A.~Kov{\'a}cs and R.~J.~M. Konings, \emph{J.~Phys.~Chem.~A}, 2011,
  \textbf{115}, 6646--6656\relax
\mciteBstWouldAddEndPuncttrue
\mciteSetBstMidEndSepPunct{\mcitedefaultmidpunct}
{\mcitedefaultendpunct}{\mcitedefaultseppunct}\relax
\EndOfBibitem
\bibitem[Kov{\'a}cs \emph{et~al.}(2012)Kov{\'a}cs, Pog{\'a}ny, and
  Konings]{Kovacs_12}
A.~Kov{\'a}cs, P.~Pog{\'a}ny and R.~J.~M. Konings, \emph{Inorg. Chem.}, 2012,
  \textbf{51}, 4841--4849\relax
\mciteBstWouldAddEndPuncttrue
\mciteSetBstMidEndSepPunct{\mcitedefaultmidpunct}
{\mcitedefaultendpunct}{\mcitedefaultseppunct}\relax
\EndOfBibitem
\bibitem[Huang \emph{et~al.}(2013)Huang, Xu, Su, Schwarz, and Li]{Huang2013}
W.~Huang, W.-H. Xu, J.~Su, W.~H.~E. Schwarz and J.~Li, \emph{Inorg. Chem.},
  2013, \textbf{52}, 14237--14245\relax
\mciteBstWouldAddEndPuncttrue
\mciteSetBstMidEndSepPunct{\mcitedefaultmidpunct}
{\mcitedefaultendpunct}{\mcitedefaultseppunct}\relax
\EndOfBibitem
\bibitem[Kov{\'a}cs \emph{et~al.}(2015)Kov{\'a}cs, Konings, Gibson, Infante,
  and Gagliardi]{Kovacs_15}
A.~Kov{\'a}cs, R.~J.~M. Konings, J.~K. Gibson, I.~Infante and L.~Gagliardi,
  \emph{Chem. Rev.}, 2015, \textbf{115}, 1725--1759\relax
\mciteBstWouldAddEndPuncttrue
\mciteSetBstMidEndSepPunct{\mcitedefaultmidpunct}
{\mcitedefaultendpunct}{\mcitedefaultseppunct}\relax
\EndOfBibitem
\bibitem[Cremer \emph{et~al.}(2002)Cremer, Filatov, Polo, Kraka, and
  Shaik]{DFT-Cremer-IJMS2002-3-604}
D.~Cremer, M.~Filatov, V.~Polo, E.~Kraka and S.~Shaik, \emph{Int. J. Mol.
  Sci.}, 2002, \textbf{3}, 604--638\relax
\mciteBstWouldAddEndPuncttrue
\mciteSetBstMidEndSepPunct{\mcitedefaultmidpunct}
{\mcitedefaultendpunct}{\mcitedefaultseppunct}\relax
\EndOfBibitem
\bibitem[Gomes \emph{et~al.}(2014)Gomes, R{\'e}al, Galland, Angeli, Cimiraglia,
  and Vallet]{actinide-Gomes-PCCP2014-16-9238}
A.~S.~P. Gomes, F.~R{\'e}al, N.~Galland, C.~Angeli, R.~Cimiraglia and
  V.~Vallet, \emph{Phys. Chem. Chem. Phys.}, 2014, \textbf{16},
  9238--9248\relax
\mciteBstWouldAddEndPuncttrue
\mciteSetBstMidEndSepPunct{\mcitedefaultmidpunct}
{\mcitedefaultendpunct}{\mcitedefaultseppunct}\relax
\EndOfBibitem
\bibitem[White(1992)]{white}
S.~R. White, \emph{Phys. Rev. Lett.}, 1992, \textbf{69}, 2863--2866\relax
\mciteBstWouldAddEndPuncttrue
\mciteSetBstMidEndSepPunct{\mcitedefaultmidpunct}
{\mcitedefaultendpunct}{\mcitedefaultseppunct}\relax
\EndOfBibitem
\bibitem[White(1993)]{white2}
S.~R. White, \emph{Phys. Rev. B}, 1993, \textbf{48}, 10345--10356\relax
\mciteBstWouldAddEndPuncttrue
\mciteSetBstMidEndSepPunct{\mcitedefaultmidpunct}
{\mcitedefaultendpunct}{\mcitedefaultseppunct}\relax
\EndOfBibitem
\bibitem[White and Martin(1999)]{white-qc}
S.~R. White and R.~L. Martin, \emph{J.~Chem.~Phys.}, 1999, \textbf{110},
  4127--4130\relax
\mciteBstWouldAddEndPuncttrue
\mciteSetBstMidEndSepPunct{\mcitedefaultmidpunct}
{\mcitedefaultendpunct}{\mcitedefaultseppunct}\relax
\EndOfBibitem
\bibitem[Schollw{\"o}ck(2005)]{scholl05}
U.~Schollw{\"o}ck, \emph{Rev. Mod. Phys.}, 2005, \textbf{77}, 259--315\relax
\mciteBstWouldAddEndPuncttrue
\mciteSetBstMidEndSepPunct{\mcitedefaultmidpunct}
{\mcitedefaultendpunct}{\mcitedefaultseppunct}\relax
\EndOfBibitem
\bibitem[Legeza \emph{et~al.}(2008)Legeza, Noack, S\'olyom, and
  Tincani]{ors_springer}
{\"O}.~Legeza, R.~Noack, J.~S\'olyom and L.~Tincani, in \emph{Computational
  Many-Particle Physics}, ed. H.~Fehske, R.~Schneider and A.~Wei\ss{}e,
  Springer, Berlin/Heidelerg, 2008, vol. 739, pp. 653--664\relax
\mciteBstWouldAddEndPuncttrue
\mciteSetBstMidEndSepPunct{\mcitedefaultmidpunct}
{\mcitedefaultendpunct}{\mcitedefaultseppunct}\relax
\EndOfBibitem
\bibitem[Marti and Reiher(2010)]{marti2010b}
K.~H. Marti and M.~Reiher, \emph{Z. Phys. Chem.}, 2010, \textbf{224},
  583--599\relax
\mciteBstWouldAddEndPuncttrue
\mciteSetBstMidEndSepPunct{\mcitedefaultmidpunct}
{\mcitedefaultendpunct}{\mcitedefaultseppunct}\relax
\EndOfBibitem
\bibitem[Chan and Sharma(2011)]{chanreview}
G.~K.-L. Chan and S.~Sharma, \emph{Annu. Rev. Phys. Chem.}, 2011, \textbf{62},
  465--481\relax
\mciteBstWouldAddEndPuncttrue
\mciteSetBstMidEndSepPunct{\mcitedefaultmidpunct}
{\mcitedefaultendpunct}{\mcitedefaultseppunct}\relax
\EndOfBibitem
\bibitem[Wouters and {Van Neck}(2014)]{wouters-review}
S.~Wouters and D.~{Van Neck}, \emph{Eur. Phys. J. D}, 2014, \textbf{68},
  272\relax
\mciteBstWouldAddEndPuncttrue
\mciteSetBstMidEndSepPunct{\mcitedefaultmidpunct}
{\mcitedefaultendpunct}{\mcitedefaultseppunct}\relax
\EndOfBibitem
\bibitem[Szalay \emph{et~al.}(2015)Szalay, Pfeffer, Murg, Barcza, Verstraete,
  Schneider, and Legeza]{Ors_ijqc}
S.~Szalay, M.~Pfeffer, V.~Murg, G.~Barcza, F.~Verstraete, R.~Schneider and
  {\"O}.~Legeza, \emph{Int.~J.~Quantum~Chem.}, 2015, \textbf{115},
  1342--1391\relax
\mciteBstWouldAddEndPuncttrue
\mciteSetBstMidEndSepPunct{\mcitedefaultmidpunct}
{\mcitedefaultendpunct}{\mcitedefaultseppunct}\relax
\EndOfBibitem
\bibitem[Yanai \emph{et~al.}(2015)Yanai, Kurashige, Mizukami, Chalupsky, Lan,
  and Saitow]{yanai-review}
T.~Yanai, Y.~Kurashige, W.~Mizukami, J.~Chalupsky, T.~N. Lan and M.~Saitow,
  \emph{Int.~J.~Quantum~Chem.}, 2015, \textbf{115}, 283--299\relax
\mciteBstWouldAddEndPuncttrue
\mciteSetBstMidEndSepPunct{\mcitedefaultmidpunct}
{\mcitedefaultendpunct}{\mcitedefaultseppunct}\relax
\EndOfBibitem
\bibitem[Legeza and S{\'o}lyom(2003)]{DEAS}
{\"O}.~Legeza and J.~S{\'o}lyom, \emph{Phys. Rev. B}, 2003, \textbf{68},
  195116\relax
\mciteBstWouldAddEndPuncttrue
\mciteSetBstMidEndSepPunct{\mcitedefaultmidpunct}
{\mcitedefaultendpunct}{\mcitedefaultseppunct}\relax
\EndOfBibitem
\bibitem[Rissler \emph{et~al.}(2006)Rissler, Noack, and White]{Rissler2006}
J.~Rissler, R.~M. Noack and S.~R. White, \emph{Chem.~Phys.}, 2006,
  \textbf{323}, 519--531\relax
\mciteBstWouldAddEndPuncttrue
\mciteSetBstMidEndSepPunct{\mcitedefaultmidpunct}
{\mcitedefaultendpunct}{\mcitedefaultseppunct}\relax
\EndOfBibitem
\bibitem[Barcza \emph{et~al.}(2011)Barcza, Legeza, Marti, and
  Reiher]{Barcza_11}
G.~Barcza, {\"O}.~Legeza, K.~H. Marti and M.~Reiher, \emph{Phys. Rev. A}, 2011,
  \textbf{83}, 012508\relax
\mciteBstWouldAddEndPuncttrue
\mciteSetBstMidEndSepPunct{\mcitedefaultmidpunct}
{\mcitedefaultendpunct}{\mcitedefaultseppunct}\relax
\EndOfBibitem
\bibitem[Legeza \emph{et~al.}(2013)Legeza, Barcza, Noack, and
  S\'o{}lyom]{Barcza2013}
{\"O}.~Legeza, G.~Barcza, R.~M. Noack and J.~S\'o{}lyom, {Entanglement topology
  of strongly correlated systems, Korrelationstage MPIPKS, Dresden}, 2013\relax
\mciteBstWouldAddEndPuncttrue
\mciteSetBstMidEndSepPunct{\mcitedefaultmidpunct}
{\mcitedefaultendpunct}{\mcitedefaultseppunct}\relax
\EndOfBibitem
\bibitem[Boguslawski \emph{et~al.}(2012)Boguslawski, Tecmer, Legeza, and
  Reiher]{entanglement_letter}
K.~Boguslawski, P.~Tecmer, {\"O}.~Legeza and M.~Reiher, \emph{J. Phys. Chem.
  Lett.}, 2012, \textbf{3}, 3129--3135\relax
\mciteBstWouldAddEndPuncttrue
\mciteSetBstMidEndSepPunct{\mcitedefaultmidpunct}
{\mcitedefaultendpunct}{\mcitedefaultseppunct}\relax
\EndOfBibitem
\bibitem[Boguslawski \emph{et~al.}(2013)Boguslawski, Tecmer, Barcza, Legeza,
  and Reiher]{entanglement_bonding_2013}
K.~Boguslawski, P.~Tecmer, G.~Barcza, {\"O}.~Legeza and M.~Reiher, \emph{J.
  Chem. Theory Comput.}, 2013, \textbf{9}, 2959--2973\relax
\mciteBstWouldAddEndPuncttrue
\mciteSetBstMidEndSepPunct{\mcitedefaultmidpunct}
{\mcitedefaultendpunct}{\mcitedefaultseppunct}\relax
\EndOfBibitem
\bibitem[Barcza \emph{et~al.}(2014)Barcza, Noack, S{\'o}lyom, and
  Legeza]{barcza2014entanglement}
G.~Barcza, R.~Noack, J.~S{\'o}lyom and {\"O}.~Legeza, \emph{Phys.~Rev.~B},
  2014, \textbf{92}, 125140\relax
\mciteBstWouldAddEndPuncttrue
\mciteSetBstMidEndSepPunct{\mcitedefaultmidpunct}
{\mcitedefaultendpunct}{\mcitedefaultseppunct}\relax
\EndOfBibitem
\bibitem[Boguslawski and Tecmer(2015)]{Kasia_ijqc}
K.~Boguslawski and P.~Tecmer, \emph{Int. J. Quantum Chem.}, 2015, \textbf{115},
  1289--1295\relax
\mciteBstWouldAddEndPuncttrue
\mciteSetBstMidEndSepPunct{\mcitedefaultmidpunct}
{\mcitedefaultendpunct}{\mcitedefaultseppunct}\relax
\EndOfBibitem
\bibitem[Keller \emph{et~al.}(2015)Keller, Boguslawski, Janowski, Reiher, and
  Pulay]{Keller2015}
S.~Keller, K.~Boguslawski, T.~Janowski, M.~Reiher and P.~Pulay, \emph{J. Chem.
  Phys.}, 2015, \textbf{142}, 244104\relax
\mciteBstWouldAddEndPuncttrue
\mciteSetBstMidEndSepPunct{\mcitedefaultmidpunct}
{\mcitedefaultendpunct}{\mcitedefaultseppunct}\relax
\EndOfBibitem
\bibitem[Boguslawski \emph{et~al.}(2016)Boguslawski, Tecmer, and
  Legeza]{Boguslawski2016}
K.~Boguslawski, P.~Tecmer and {\"O}.~Legeza, \emph{arXiv:1606.08503
  [cond-mat.str-el]}, 2016,  1--15\relax
\mciteBstWouldAddEndPuncttrue
\mciteSetBstMidEndSepPunct{\mcitedefaultmidpunct}
{\mcitedefaultendpunct}{\mcitedefaultseppunct}\relax
\EndOfBibitem
\bibitem[Mottet \emph{et~al.}(2014)Mottet, Tecmer, Boguslawski, Legeza, and
  Reiher]{PCCP_bonding}
M.~Mottet, P.~Tecmer, K.~Boguslawski, {\"O}.~Legeza and M.~Reiher, \emph{Phys.
  Chem. Chem. Phys.}, 2014, \textbf{16}, 8872--8880\relax
\mciteBstWouldAddEndPuncttrue
\mciteSetBstMidEndSepPunct{\mcitedefaultmidpunct}
{\mcitedefaultendpunct}{\mcitedefaultseppunct}\relax
\EndOfBibitem
\bibitem[Szilv{\'a}si \emph{et~al.}(2015)Szilv{\'a}si, Barcza, and
  Legeza]{bonding_qit}
T.~Szilv{\'a}si, G.~Barcza and {\"O}.~Legeza, \emph{arXiv:1509.04241
  [physics.chem-ph]}, 2015,  1--7\relax
\mciteBstWouldAddEndPuncttrue
\mciteSetBstMidEndSepPunct{\mcitedefaultmidpunct}
{\mcitedefaultendpunct}{\mcitedefaultseppunct}\relax
\EndOfBibitem
\bibitem[Boguslawski and Reiher(2014)]{boguslawski2014chemical}
K.~Boguslawski and M.~Reiher, in \emph{The Chemical Bond: Chemical Bonding
  Across the Periodic Table}, ed. G.~Frenking and S.~Shaik, Wiley-VCH Verlag
  GmbH \& Co. KGaA, 2014, ch.~8, pp. 219--252\relax
\mciteBstWouldAddEndPuncttrue
\mciteSetBstMidEndSepPunct{\mcitedefaultmidpunct}
{\mcitedefaultendpunct}{\mcitedefaultseppunct}\relax
\EndOfBibitem
\bibitem[Freitag \emph{et~al.}(2015)Freitag, Knecht, Keller, Delcey, Aquilante,
  Pedersen, Lindh, Reiher, and Gonzalez]{Roland-RuNO}
L.~Freitag, S.~Knecht, S.~F. Keller, M.~G. Delcey, F.~Aquilante, T.~B.
  Pedersen, R.~Lindh, M.~Reiher and L.~Gonzalez, \emph{Phys. Chem. Chem.
  Phys.}, 2015, \textbf{17}, 13769--13769\relax
\mciteBstWouldAddEndPuncttrue
\mciteSetBstMidEndSepPunct{\mcitedefaultmidpunct}
{\mcitedefaultendpunct}{\mcitedefaultseppunct}\relax
\EndOfBibitem
\bibitem[Tecmer \emph{et~al.}(2015)Tecmer, Boguslawski, and
  Ayers]{AP1roG-actinides}
P.~Tecmer, K.~Boguslawski and P.~Ayers, \emph{Phys. Chem. Chem. Phys.}, 2015,
  \textbf{17}, 14427--14436\relax
\mciteBstWouldAddEndPuncttrue
\mciteSetBstMidEndSepPunct{\mcitedefaultmidpunct}
{\mcitedefaultendpunct}{\mcitedefaultseppunct}\relax
\EndOfBibitem
\bibitem[Zhao \emph{et~al.}(2015)Zhao, Boguslawski, Tecmer, Duperrouzel,
  Barcza, Legeza, and Ayers]{Zhao2015}
Y.~Zhao, K.~Boguslawski, P.~Tecmer, C.~Duperrouzel, G.~Barcza, {\"{O}}.~Legeza
  and P.~W. Ayers, \emph{Theor. Chem. Acc.}, 2015, \textbf{134}, 120\relax
\mciteBstWouldAddEndPuncttrue
\mciteSetBstMidEndSepPunct{\mcitedefaultmidpunct}
{\mcitedefaultendpunct}{\mcitedefaultseppunct}\relax
\EndOfBibitem
\bibitem[Murg \emph{et~al.}(2015)Murg, Verstraete, Schneider, Nagy, and
  Legeza]{Ors-LiF-TTNS}
V.~Murg, F.~Verstraete, R.~Schneider, P.~R. Nagy and {\"O}.~Legeza,
  \emph{J.~Chem.~Theory~Comput.}, 2015, \textbf{11}, 1027--1036\relax
\mciteBstWouldAddEndPuncttrue
\mciteSetBstMidEndSepPunct{\mcitedefaultmidpunct}
{\mcitedefaultendpunct}{\mcitedefaultseppunct}\relax
\EndOfBibitem
\bibitem[Fertitta \emph{et~al.}(2014)Fertitta, Paulus, Barcza, and
  Legeza]{MIT-Fertita-2014}
E.~Fertitta, B.~Paulus, G.~Barcza and {\"O}.~Legeza, \emph{Phys. Rev. B}, 2014,
  \textbf{90}, 245129\relax
\mciteBstWouldAddEndPuncttrue
\mciteSetBstMidEndSepPunct{\mcitedefaultmidpunct}
{\mcitedefaultendpunct}{\mcitedefaultseppunct}\relax
\EndOfBibitem
\bibitem[Duperrouzel \emph{et~al.}(2015)Duperrouzel, Tecmer, Boguslawski,
  Barcza, Legeza, and Ayers]{Corinne_2015}
C.~Duperrouzel, P.~Tecmer, K.~Boguslawski, G.~Barcza, {\"O}.~Legeza and P.~W.
  Ayers, \emph{Chem. Phys. Lett.}, 2015, \textbf{621}, 160--164\relax
\mciteBstWouldAddEndPuncttrue
\mciteSetBstMidEndSepPunct{\mcitedefaultmidpunct}
{\mcitedefaultendpunct}{\mcitedefaultseppunct}\relax
\EndOfBibitem
\bibitem[Tecmer \emph{et~al.}(2014)Tecmer, Boguslawski, Legeza, and
  Reiher]{CUO_DMRG}
P.~Tecmer, K.~Boguslawski, {\"O}.~Legeza and M.~Reiher, \emph{Phys. Chem. Chem.
  Phys}, 2014, \textbf{16}, 719--727\relax
\mciteBstWouldAddEndPuncttrue
\mciteSetBstMidEndSepPunct{\mcitedefaultmidpunct}
{\mcitedefaultendpunct}{\mcitedefaultseppunct}\relax
\EndOfBibitem
\bibitem[Maron \emph{et~al.}(1999)Maron, Leininger, Schimmelpfennig, Vallet,
  Heully, Teichteil, Gropen, and Wahlgren]{actinide-Maron-CP1999-244-195}
L.~Maron, T.~Leininger, B.~Schimmelpfennig, V.~Vallet, J.-L. Heully,
  C.~Teichteil, O.~Gropen and U.~Wahlgren, \emph{Chem. Phys.}, 1999,
  \textbf{244}, 195--201\relax
\mciteBstWouldAddEndPuncttrue
\mciteSetBstMidEndSepPunct{\mcitedefaultmidpunct}
{\mcitedefaultendpunct}{\mcitedefaultseppunct}\relax
\EndOfBibitem
\bibitem[Infante \emph{et~al.}(2006)Infante, Gomes, and Visscher]{fscc-npo2}
I.~Infante, A.~S.~P. Gomes and L.~Visscher, \emph{J.~Chem.~Phys.}, 2006,
  \textbf{125}, 074301\relax
\mciteBstWouldAddEndPuncttrue
\mciteSetBstMidEndSepPunct{\mcitedefaultmidpunct}
{\mcitedefaultendpunct}{\mcitedefaultseppunct}\relax
\EndOfBibitem
\bibitem[Macchia \emph{et~al.}(2008)Macchia, Infante, Raab, Gibson, and
  Gagliardi]{ivan_puo2_08}
G.~L. Macchia, I.~Infante, J.~Raab, J.~K. Gibson and L.~Gagliardi,
  \emph{Phys.~Chem.~Chem.~Phys.}, 2008, \textbf{48}, 7278--7283\relax
\mciteBstWouldAddEndPuncttrue
\mciteSetBstMidEndSepPunct{\mcitedefaultmidpunct}
{\mcitedefaultendpunct}{\mcitedefaultseppunct}\relax
\EndOfBibitem
\bibitem[Denning(2007)]{denning07}
R.~G. Denning, \emph{J.~Phys.~Chem.~A}, 2007, \textbf{111}, 4125--4143\relax
\mciteBstWouldAddEndPuncttrue
\mciteSetBstMidEndSepPunct{\mcitedefaultmidpunct}
{\mcitedefaultendpunct}{\mcitedefaultseppunct}\relax
\EndOfBibitem
\bibitem[Morss \emph{et~al.}(2011)Morss, Edelstein, Fuger, and
  Katz]{Actinides_bible}
\emph{The Chemistry of the Actinide and Transactinide Elements}, ed. L.~R.
  Morss, N.~M. Edelstein, J.~Fuger and J.~J. Katz, Springer, Dordrecht, The
  Netherlands, 4th edn., 2011\relax
\mciteBstWouldAddEndPuncttrue
\mciteSetBstMidEndSepPunct{\mcitedefaultmidpunct}
{\mcitedefaultendpunct}{\mcitedefaultseppunct}\relax
\EndOfBibitem
\bibitem[Wang \emph{et~al.}(2012)Wang, van Gunsteren, and
  Chai]{actinoid_rev_2012}
D.~Wang, W.~F. van Gunsteren and Z.~Chai, \emph{Chem. Soc. Rev.}, 2012,
  \textbf{41}, 5836--5865\relax
\mciteBstWouldAddEndPuncttrue
\mciteSetBstMidEndSepPunct{\mcitedefaultmidpunct}
{\mcitedefaultendpunct}{\mcitedefaultseppunct}\relax
\EndOfBibitem
\bibitem[Denning(1992)]{denning_91b}
R.~G. Denning, \emph{Struct. Bonding}, 1992, \textbf{79}, 215--276\relax
\mciteBstWouldAddEndPuncttrue
\mciteSetBstMidEndSepPunct{\mcitedefaultmidpunct}
{\mcitedefaultendpunct}{\mcitedefaultseppunct}\relax
\EndOfBibitem
\bibitem[Castro-Rodriguez \emph{et~al.}(2004)Castro-Rodriguez, Nakai, Zakharov,
  Rheingold, and Meyer]{Castro-Rodriguez2004}
I.~Castro-Rodriguez, H.~Nakai, L.~N. Zakharov, A.~L. Rheingold and K.~Meyer,
  \emph{Science}, 2004, \textbf{305}, 1757--1759\relax
\mciteBstWouldAddEndPuncttrue
\mciteSetBstMidEndSepPunct{\mcitedefaultmidpunct}
{\mcitedefaultendpunct}{\mcitedefaultseppunct}\relax
\EndOfBibitem
\bibitem[Baker(2012)]{Baker2012}
R.~J. Baker, \emph{Chem. Eur. J.}, 2012, \textbf{18}, 16258--16271\relax
\mciteBstWouldAddEndPuncttrue
\mciteSetBstMidEndSepPunct{\mcitedefaultmidpunct}
{\mcitedefaultendpunct}{\mcitedefaultseppunct}\relax
\EndOfBibitem
\bibitem[Zaitsevskii \emph{et~al.}(2013)Zaitsevskii, Titov, Mal'kov, Tananaev,
  and Kiselev]{actinide-Zaitsevskii-DC2013-448-1}
A.~V. Zaitsevskii, A.~V. Titov, S.~S. Mal'kov, I.~G. Tananaev and Y.~M.
  Kiselev, \emph{Dokl. Chem.}, 2013, \textbf{448}, 1--3\relax
\mciteBstWouldAddEndPuncttrue
\mciteSetBstMidEndSepPunct{\mcitedefaultmidpunct}
{\mcitedefaultendpunct}{\mcitedefaultseppunct}\relax
\EndOfBibitem
\bibitem[Gao \emph{et~al.}(2004)Gao, Zhu, Wang, Sun, and
  Meng]{actinide-Gao-AC2004-62-454}
T.~Gao, Z.-H. Zhu, X.-L. Wang, Y.~Sun and D.-Q. Meng, \emph{Acta Chim}, 2004,
  \textbf{62}, 454--460\relax
\mciteBstWouldAddEndPuncttrue
\mciteSetBstMidEndSepPunct{\mcitedefaultmidpunct}
{\mcitedefaultendpunct}{\mcitedefaultseppunct}\relax
\EndOfBibitem
\bibitem[Andrae \emph{et~al.}(1990)Andrae, H{\"a}u{\ss}ermann, Dolg, Stoll, and
  Preu{\ss}]{ECP-Andrae-TCA1990-77-123--141}
D.~Andrae, U.~H{\"a}u{\ss}ermann, M.~Dolg, H.~Stoll and H.~Preu{\ss},
  \emph{Theo. Chem. Acta}, 1990, \textbf{77}, 123--141\relax
\mciteBstWouldAddEndPuncttrue
\mciteSetBstMidEndSepPunct{\mcitedefaultmidpunct}
{\mcitedefaultendpunct}{\mcitedefaultseppunct}\relax
\EndOfBibitem
\bibitem[Cao \emph{et~al.}(2003)Cao, Dolg, and
  Stoll]{basis-Cao-JCP2003-118-487-496}
X.~Cao, M.~Dolg and H.~Stoll, \emph{J.~Chem.~Phys.}, 2003, \textbf{118},
  487--496\relax
\mciteBstWouldAddEndPuncttrue
\mciteSetBstMidEndSepPunct{\mcitedefaultmidpunct}
{\mcitedefaultendpunct}{\mcitedefaultseppunct}\relax
\EndOfBibitem
\bibitem[Cao and Dolg(2004)]{basis-Cao-JMST2004-673-203---209}
X.~Cao and M.~Dolg, \emph{J. Mol. Struct. \{THEOCHEM\}}, 2004, \textbf{673},
  203 -- 209\relax
\mciteBstWouldAddEndPuncttrue
\mciteSetBstMidEndSepPunct{\mcitedefaultmidpunct}
{\mcitedefaultendpunct}{\mcitedefaultseppunct}\relax
\EndOfBibitem
\bibitem[Dunning(1989)]{basis-Dunning-JCP1989-90-1007}
T.~H. Dunning, Jr., \emph{J. Chem. Phys.}, 1989, \textbf{90}, 1007--1023\relax
\mciteBstWouldAddEndPuncttrue
\mciteSetBstMidEndSepPunct{\mcitedefaultmidpunct}
{\mcitedefaultendpunct}{\mcitedefaultseppunct}\relax
\EndOfBibitem
\bibitem[Kendall \emph{et~al.}(1992)Kendall, Dunning, and
  Harrison]{basis-Kendall-JCP1992-96-6796}
R.~A. Kendall, T.~H. Dunning, Jr. and R.~J. Harrison, \emph{J. Chem. Phys.},
  1992, \textbf{96}, 6796--6806\relax
\mciteBstWouldAddEndPuncttrue
\mciteSetBstMidEndSepPunct{\mcitedefaultmidpunct}
{\mcitedefaultendpunct}{\mcitedefaultseppunct}\relax
\EndOfBibitem
\bibitem[Frisch \emph{et~al.}()Frisch, Trucks, Schlegel, Scuseria, Robb,
  Cheeseman, Montgomery, Vreven, Kudin, Burant, Millam, Iyengar, Tomasi,
  Barone, Mennucci, Cossi, Scalmani, Rega, Petersson, Nakatsuji, Hada, Ehara,
  Toyota, Fukuda, Hasegawa, Ishida, Nakajima, Honda, Kitao, Nakai, Klene, Li,
  Knox, Hratchian, Cross, Bakken, Adamo, Jaramillo, Gomperts, Stratmann,
  Yazyev, Austin, Cammi, Pomelli, Ochterski, Ayala, Morokuma, Voth, Salvador,
  Dannenberg, Zakrzewski, Dapprich, Daniels, Strain, Farkas, Malick, Rabuck,
  Raghavachari, Foresman, Ortiz, Cui, Baboul, Clifford, Cioslowski, Stefanov,
  Liu, Liashenko, Piskorz, Komaromi, Martin, Fox, Keith, Al-Laham, Peng,
  Nanayakkara, Challacombe, Gill, Johnson, Chen, Wong, Gonzalez, and
  Pople]{prog-G09}
M.~J. Frisch, G.~W. Trucks, H.~B. Schlegel, G.~E. Scuseria, M.~A. Robb, J.~R.
  Cheeseman, J.~A. Montgomery, Jr., T.~Vreven, K.~N. Kudin, J.~C. Burant, J.~M.
  Millam, S.~S. Iyengar, J.~Tomasi, V.~Barone, B.~Mennucci, M.~Cossi,
  G.~Scalmani, N.~Rega, G.~A. Petersson, H.~Nakatsuji, M.~Hada, M.~Ehara,
  K.~Toyota, R.~Fukuda, J.~Hasegawa, M.~Ishida, T.~Nakajima, Y.~Honda,
  O.~Kitao, H.~Nakai, M.~Klene, X.~Li, J.~E. Knox, H.~P. Hratchian, J.~B.
  Cross, V.~Bakken, C.~Adamo, J.~Jaramillo, R.~Gomperts, R.~E. Stratmann,
  O.~Yazyev, A.~J. Austin, R.~Cammi, C.~Pomelli, J.~W. Ochterski, P.~Y. Ayala,
  K.~Morokuma, G.~A. Voth, P.~Salvador, J.~J. Dannenberg, V.~G. Zakrzewski,
  S.~Dapprich, A.~D. Daniels, M.~C. Strain, O.~Farkas, D.~K. Malick, A.~D.
  Rabuck, K.~Raghavachari, J.~B. Foresman, J.~V. Ortiz, Q.~Cui, A.~G. Baboul,
  S.~Clifford, J.~Cioslowski, B.~B. Stefanov, G.~Liu, A.~Liashenko, P.~Piskorz,
  I.~Komaromi, R.~L. Martin, D.~J. Fox, T.~Keith, M.~A. Al-Laham, C.~Y. Peng,
  A.~Nanayakkara, M.~Challacombe, P.~M.~W. Gill, B.~Johnson, W.~Chen, M.~W.
  Wong, C.~Gonzalez and J.~A. Pople, \emph{Gaussian 09, \uppercase{R}evision
  \uppercase{C}.02}, \uppercase{G}aussian, Inc., Wallingford, CT, 2004\relax
\mciteBstWouldAddEndPuncttrue
\mciteSetBstMidEndSepPunct{\mcitedefaultmidpunct}
{\mcitedefaultendpunct}{\mcitedefaultseppunct}\relax
\EndOfBibitem
\bibitem[Becke(1993)]{dft-Becke-JCP1993-98-5648-5652}
A.~D. Becke, \emph{J.~Chem.~Phys.}, 1993, \textbf{98}, 5648--5652\relax
\mciteBstWouldAddEndPuncttrue
\mciteSetBstMidEndSepPunct{\mcitedefaultmidpunct}
{\mcitedefaultendpunct}{\mcitedefaultseppunct}\relax
\EndOfBibitem
\bibitem[Roos \emph{et~al.}(1980)Roos, Taylor, and Siegbahn]{Roos_casscf}
B.~O. Roos, P.~R. Taylor and P.~E.~M. Siegbahn, \emph{Chem. Phys.}, 1980,
  \textbf{48}, 157--173\relax
\mciteBstWouldAddEndPuncttrue
\mciteSetBstMidEndSepPunct{\mcitedefaultmidpunct}
{\mcitedefaultendpunct}{\mcitedefaultseppunct}\relax
\EndOfBibitem
\bibitem[Werner and Knowles(1985)]{Werner_1985}
H.-J. Werner and P.~J. Knowles, \emph{J. Chem. Phys.}, 1985, \textbf{82},
  5053--5063\relax
\mciteBstWouldAddEndPuncttrue
\mciteSetBstMidEndSepPunct{\mcitedefaultmidpunct}
{\mcitedefaultendpunct}{\mcitedefaultseppunct}\relax
\EndOfBibitem
\bibitem[{Werner} \emph{et~al.}(2012){Werner}, {Knowles}, Lindh, Manby,
  M.~Sch{\"ui}tz, Korona, Mitrushenkov, Rauhut, Adler, Amos, Bernhardsson,
  Berning, Cooper, Deegan, Dobbyn, Eckert, Goll, Hampel, Hetzer, Hrenar,
  Knizia, K{\"o}ppl, Liu, Lloyd, Mata, May, McNicholas, Meyer, Mura, Nicklass,
  Palmieri, Pfl{\"u}ger, Pitzer, Reiher, Schumann, Stoll, Stone, Tarroni,
  Thorsteinsson, Wang, and Wolf]{molpro2012}
H.-J. {Werner}, P.~J. {Knowles}, R.~Lindh, F.~R. Manby, P.~C. M.~Sch{\"ui}tz,
  T.~Korona, A.~Mitrushenkov, G.~Rauhut, T.~B. Adler, R.~D. Amos,
  A.~Bernhardsson, A.~Berning, D.~L. Cooper, M.~J.~O. Deegan, A.~J. Dobbyn,
  F.~Eckert, E.~Goll, C.~Hampel, G.~Hetzer, T.~Hrenar, G.~Knizia, C.~K{\"o}ppl,
  Y.~Liu, A.~W. Lloyd, R.~A. Mata, A.~J. May, S.~J. McNicholas, W.~Meyer, M.~E.
  Mura, A.~Nicklass, P.~Palmieri, K.~Pfl{\"u}ger, R.~Pitzer, M.~Reiher,
  U.~Schumann, H.~Stoll, A.~J. Stone, R.~Tarroni, T.~Thorsteinsson, M.~Wang and
  A.~Wolf, \emph{MOLPRO, Version 2012.1, A Package Of \emph{Ab Initio}
  Programs}, 2012, see http://www.molpro.net\relax
\mciteBstWouldAddEndPuncttrue
\mciteSetBstMidEndSepPunct{\mcitedefaultmidpunct}
{\mcitedefaultendpunct}{\mcitedefaultseppunct}\relax
\EndOfBibitem
\bibitem[Werner \emph{et~al.}(2012)Werner, Knowles, Knizia, Manby, and
  Sch{\"u}tz]{molpro-WIREs}
H.-J. Werner, P.~J. Knowles, G.~Knizia, F.~R. Manby and M.~Sch{\"u}tz,
  \emph{WIREs Comput. Mol. Sci.}, 2012, \textbf{2}, 242--253\relax
\mciteBstWouldAddEndPuncttrue
\mciteSetBstMidEndSepPunct{\mcitedefaultmidpunct}
{\mcitedefaultendpunct}{\mcitedefaultseppunct}\relax
\EndOfBibitem
\bibitem[Jmo()]{Jmol}
Jmol: An Open-Source Java Viewer for Chemical Structures in 3D. {\tt
  http://www.jmol.org/}\relax
\mciteBstWouldAddEndPuncttrue
\mciteSetBstMidEndSepPunct{\mcitedefaultmidpunct}
{\mcitedefaultendpunct}{\mcitedefaultseppunct}\relax
\EndOfBibitem
\bibitem[Legeza()]{dmrg_ors}
{\"O}.~Legeza, \emph{\textsc{QC-DMRG-Budapest}, A Program for Quantum Chemical
  {DMRG} Calculations. { \rm Copyright 2000--2016, HAS RISSPO Budapest}}\relax
\mciteBstWouldAddEndPuncttrue
\mciteSetBstMidEndSepPunct{\mcitedefaultmidpunct}
{\mcitedefaultendpunct}{\mcitedefaultseppunct}\relax
\EndOfBibitem
\bibitem[Legeza \emph{et~al.}(2003)Legeza, R{\"o}der, and Hess]{legeza2003}
{\"O}.~Legeza, J.~R{\"o}der and B.~A. Hess, \emph{Phys.~Rev.~B}, 2003,
  \textbf{67}, 125114\relax
\mciteBstWouldAddEndPuncttrue
\mciteSetBstMidEndSepPunct{\mcitedefaultmidpunct}
{\mcitedefaultendpunct}{\mcitedefaultseppunct}\relax
\EndOfBibitem
\bibitem[Legeza and S{\'o}lyom(2004)]{legeza2004}
{\"O}.~Legeza and J.~S{\'o}lyom, \emph{Phys.~Rev.~B}, 2004, \textbf{70},
  205118\relax
\mciteBstWouldAddEndPuncttrue
\mciteSetBstMidEndSepPunct{\mcitedefaultmidpunct}
{\mcitedefaultendpunct}{\mcitedefaultseppunct}\relax
\EndOfBibitem
\bibitem[Legeza and S\'{o}lyom(2006)]{Legeza2006}
{\"O}.~Legeza and J.~S\'{o}lyom, \emph{Phys.~Rev.~Lett.}, 2006, \textbf{96},
  4--7\relax
\mciteBstWouldAddEndPuncttrue
\mciteSetBstMidEndSepPunct{\mcitedefaultmidpunct}
{\mcitedefaultendpunct}{\mcitedefaultseppunct}\relax
\EndOfBibitem
\bibitem[Stein and Reiher(2016)]{Stein2016}
C.~J. Stein and M.~Reiher, \emph{J.~Chem.~Theory~Comput.}, 2016, \textbf{12},
  1760--1771\relax
\mciteBstWouldAddEndPuncttrue
\mciteSetBstMidEndSepPunct{\mcitedefaultmidpunct}
{\mcitedefaultendpunct}{\mcitedefaultseppunct}\relax
\EndOfBibitem
\bibitem[Szalay \emph{et~al.}(2016)Szalay, Barcza, Szilv\'asi, Veis, and
  Legeza]{Szalay2016}
S.~Szalay, G.~Barcza, T.~Szilv\'asi, L.~Veis and {\"O}.~Legeza,
  \emph{arXiv:1605.06919 [quant-ph]}, 2016,  1--10\relax
\mciteBstWouldAddEndPuncttrue
\mciteSetBstMidEndSepPunct{\mcitedefaultmidpunct}
{\mcitedefaultendpunct}{\mcitedefaultseppunct}\relax
\EndOfBibitem
\end{mcitethebibliography}
\bibliographystyle{rsc} 

\end{document}